\documentclass[12pt]{iopart}
\usepackage{iopams}
\usepackage{verbatim}
\usepackage{graphicx}
\usepackage{bm}
\usepackage{color}
\usepackage{hyperref}
\usepackage{ulem}
\usepackage{multirow}
\usepackage{braket}
\newcommand{\naiseki}[2]{\left\langle #1 | #2 \right\rangle}
\newcommand{\identity}{\mbox{1}\hspace{-0.25em}\mbox{l}}

\begin{document}

\title{Stability of topologically protected edge states in nonlinear quantum walks: Additional bifurcations unique to Floquet systems}

\author{Ken Mochizuki$^1$, Norio Kawakami$^2$, and Hideaki Obuse$^1$}
\address{$^1$Department of Applied Physics, Hokkaido University, Sapporo 060-8628, Japan}
\address{$^2$Department of Physics, Kyoto University, Kyoto 606-8502, Japan}
\ead{ken\_mochizuki@eng.hokudai.ac.jp}
\vspace{10pt}

\begin{abstract}
Recently, effects of nonlinearity on topologically nontrivial systems have attracted attention and the stability of topologically protected edge states has been studied for a quantum walk with nonlinear effects, which is akin to time-periodically driven systems (Floquet systems). In the previous work, it has been found that the edge states can be stable attractors or unstable repellers depending on their intrinsic topological property, while the stability is not affected by the strength of nonlinearity. In the present work, we find additional bifurcations at which edge states change from stable attractors to unstable repellers with increasing the strength of nonlinearity in nonlinear quantum walks, for the first time. The new bifurcations are unique to Floquet systems, since we take dynamical properties of Floquet systems into consideration by directly applying the time-evolution operator of the quantum walks to the linear stability analysis. Our results shed new light on nonlinear effects 
 on topological edge states in Floquet systems.
\end{abstract}

\section{Introduction}
\label{sec:introduction}
The study of topological phases of matter has been very active in broad research fields, from original condensed matter physics to photonic, acoustic, and exitonic systems, and so on. In the latter new systems, the research of nonlinear effects has attracted theoretical and experimental interests \cite{bardyn2016chiral,leykam2016edge,kartashov2017bistable,dobrykh2018nonlinear,poddubny2018ring,
malzard2018nonlinear,kruk2019nonlinear} since nonlinear effects can cause interesting phenomena related to topologically protected edge states, such as emergence of solitons \cite{leykam2016edge,poddubny2018ring}, frequency shift of edge modes \cite{dobrykh2018nonlinear}, and non-reciprocity \cite{kruk2019nonlinear}. While the nonlinear effects have been actively studied for topological phases in static systems, those for topological phases in time-dependent systems have been unclear, because of difficulty for treating nonlinear effects in time-dependent systems, even in periodically driven systems.\\\indent
Floquet systems, time-periodically driven systems, have attracted great deal of attention \cite{bukov2015universal,oka2019floquet} to explore non-trivial topological phases induced by periodically driving external fields \cite{oka2009photovoltaic,kitagawa2010topological,kitagawa2011transport,lindner2011floquet,
jiang2011majorana,wang2013observation,gometzleon2013floquet,calvo2015floquet,roy2017floquet,
morimoto2017floquet,hockendorf2019universal}, analyze the stability of systems itself or limit cycles \cite{gokccek2004stability,klausmeier2008floquet,boland2009limit,bao2010complex,hasegawa2013enhanced,
seiwert2013instability,milicevic2014experimental,mari2014floquet,marconi2015vectorial,chan2015limit,
hu2017exponential}, to name a few.
Floquet systems with high tunability have been realized in various experimental setups. One example is a discrete time quantum walk (hereafter, quantum walk)  \cite{karski2009quantum,schmitz2009quantum,broome2010discrete,schreiber2010photons,
zahringer2010realization,regensburger2011photon,kitagawa2012observation,
regensburger2012parity,sansoni2012two,schreiber20122d,crespi2013anderson,genske2013electric,
cardano2015quantum,zhao2015experimental,xiao2017observation,chen2018observation}. 
Quantum walks are described by time-evolution operators and have been exploited to explore Floquet topological phases \cite{kitagawa2012observation,xiao2017observation,chen2018observation,kitagawa2010exploring,
obuse2011topological,asboth2012symmetries,asboth2013bulk,asboth2014chiral,obuse2015unveiling,
cardano2016statistical,zhang2017decomposition,barkhofen2018supersymmetric,kawasaki2019bulk}.
There are several advantages in quantum walks in comparison to condensed matter systems, such as real space observations of edge states \cite{kitagawa2012observation} and existence of edge states with long lifetime in the presence of dissipation  \cite{xiao2017observation}. In addition to these unique features, nonlinear effects in quantum walks using a feed-forward scheme have been studied \cite{shikano2014discrete,lee2015quantum}.\\\indent
In 2016, it was shown that, in a nonlinear quamtum walk, topologically protected edge states become a stable attractor or an unstable repeller after long time evolution, depending only on their topological properties \cite{gerasimenko2016attractor}, but not on the strength of nonlinearity. To this end, the linear stability analysis was applied using an effective Hamiltonian in a continuum limit. However, taking the continuum limit results in losing an exact description of dynamical properties of the system, which plays a key role in Floquet systems.\\\indent
In this work, we study the stability of edge states in time-periodically driven systems with nonlinear effects, by focusing on nonlinear quantum walks. To this end, we apply the linear stability analysis of edge states in nonlinear quantum walks, by directly treating time-evolution operators to take the dynamical properties into account. As a consequence, in addition to obtaining consistent results in Ref.  \cite{gerasimenko2016attractor}, we find additional bifurcations where edge states change from stable attractors to unstable repellers with increasing the strength of nonlinearity. The bifurcations are typical examples of phenomena unique to Floquet systems, and cannot be predicted by the approach of the effective Hamiltonians in the continuum limit. We discuss the origin of this bifurcation at the end of this paper.\\\indent
This paper is organized as follows. In Sec. \ref{sec:linear}, we explain linear quantum walks and edge states. The stability analysis done in Ref.  \cite{gerasimenko2016attractor} is reviewed in Sec. \ref{sec:previous}. Section \ref{sec:present} is devoted to a linear stability analysis for edge states of single (\ref{subsec:1step}) and two (\ref{subsec:2step}) step nonlinear quantum walks, by directly treating time-evolution operators. Summary and discussion are given in Sec. \ref{sec:discussion}.

\section{Quantum walks without nonlinear effects}
\label{sec:linear}
Before dealing with nonlinear quantum walks, we explain standard quantum walks without nonlinearity. We consider quantum walks in which walkers move in one dimensional position space $\ket{x}$ and have two internal states $\ket{L}=(1,0)^{\rm T}$ and $\ket{R}=(0,1)^{\rm T}$, where the superscript T denotes the transpose. Using these bases, a state in a time step $t$ is written as
\begin{equation}
\ket{\psi(t)}
=\sum_{x,s=L,R}\psi_{x,s}(t)
\ket{x}\otimes\ket{s},
\label{eq:psi}
\end{equation}
where $\psi_{x,s}(t)$ denotes the wave function amplitude. By using a time-evolution operator $U$, the state at time step $t+1$ is described as
\begin{equation}
\ket{\psi(t+1)}=U\ket{\psi(t)}.
\label{eq:time_evolution_linear}
\end{equation}
\begin{figure}[b]
\begin{center}
\includegraphics[scale=0.30]{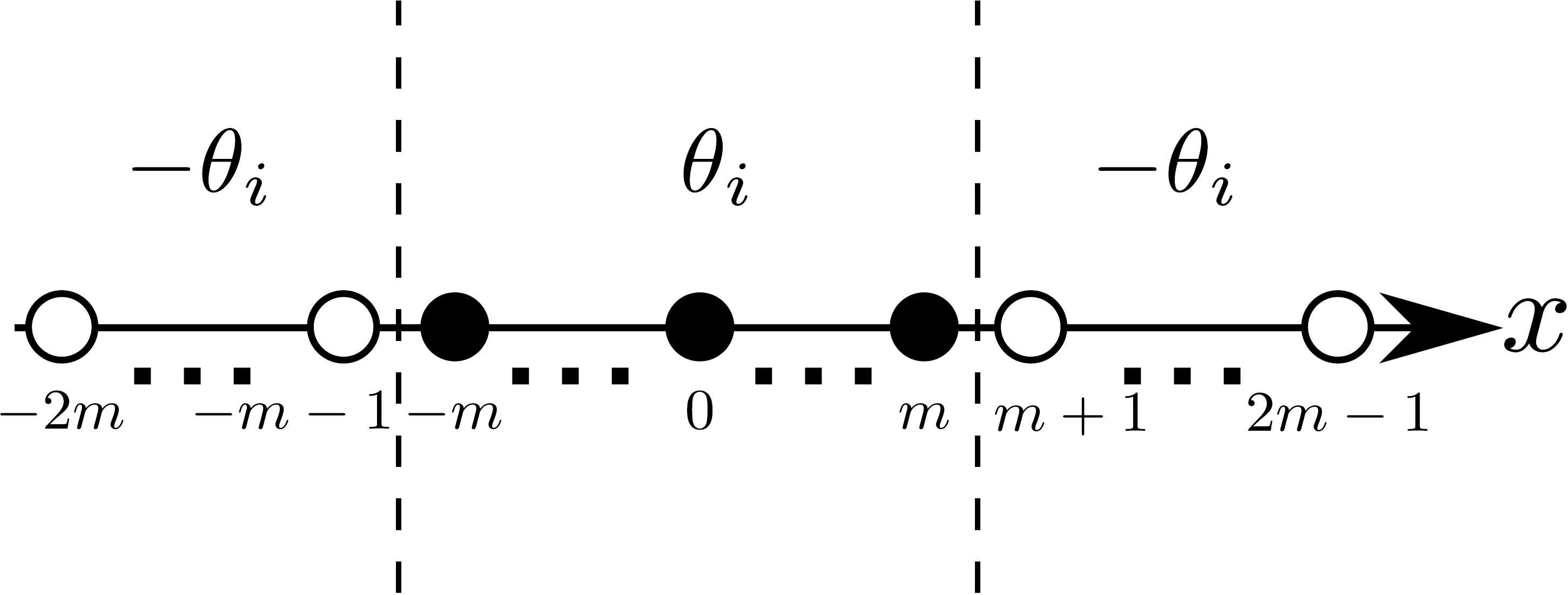}
\caption{Position dependence of $\theta_i(x)\ (i=0,1,2)$. There are two
 boundaries between $x=\pm m$ and $x=\pm(m+1)$. The number of sites is
 $4m$ and $m$ is an even number. We call the region for $|x|\le m$
 ($|x|>m$) as the inner (outer) region. Periodic boundary conditions are
 imposed on both edges at $x=-2m$ and $x=2m-1$.}
\label{fig:theta}
\end{center}
\end{figure}
In the standard quantum walks, $U$ is composed of a coin operator $C(\theta)$ and a shift operator $S$, which are defined as
\begin{equation}
C(\theta)=\sum_x\ket{x}\bra{x}
\otimes{\mathcal C}[\theta(x)],\ \ 
{\mathcal C}[\theta(x)]=
e^{-i\theta(x)\sigma_2},
\label{eq:coin_operator}
\end{equation}
\begin{equation}
S=\sum_x\left(\begin{array}{cc}
\ket{x-1}\bra{x}&0\\
0&\ket{x+1}\bra{x}
\end{array}\right),
\label{eq:shift_operator}
\end{equation}
respectively, where $\sigma_2$ is one of Pauli matrices $\sigma_i\ (i=1,2,3)$. The position-dependent rotation angle $\theta(x)$ is a parameter which  determines how internal states are mixed, and the shift operator changes the position of wave function amplitudes depending on internal states. In quantum walks, quasienergy $\varepsilon$ is defined as $\mu=e^{-i\varepsilon}$, where $\mu$ is the eigenvalue of $U$. The quasienergy $\varepsilon$ has $2\pi$ periodicity.

We consider two types of quantum walks. The first one is a single-step quantum walk, whose time-evolution operator is given by
\begin{equation}
U_1=C(\theta_0/2)SC(\theta_0/2).
\label{eq:time_evolution_operator_1step}
\end{equation}
The second one is a two-step quantum walk defined by
\begin{equation}
U_2=U_{2b}U_{2a},
\label{eq:time_evolution_operator_2step}
\end{equation}
where $U_{2a}$ and $U_{2b}$ are
\begin{equation}
U_{2a}=C(\theta_2/2)SC(\theta_1/2),\ 
U_{2b}=C(\theta_1/2)SC(\theta_2/2).
\label{eq:time_evolution_operator_2step_ab}
\end{equation}
Note that, since $U_2$ shifts wave function amplitudes on even (odd) sites to even (odd) sites as it includes two shift operators, $U_2$ is decomposed to a block matrix structure for even sites and odd sites. As defined in a symmetry time frame \cite{asboth2012symmetries}, time-evolution operators of single-step and two-step quantum walks, $U_1$ and $U_2$, have time-reversal symmetry $\mathcal{T}U\mathcal{T}^{-1}=U^{-1}$, particle-hole symmetry $\Xi U \Xi^{-1}=U$, and chiral symmetry $\Gamma U \Gamma^{-1}=U^{-1}$. Each symmetry operator is defined as  $\mathcal{T}=\identity_x\otimes\sigma_1\mathcal{K},\ \Xi=\identity_x\otimes\identity_s\mathcal{K},$ and $\Gamma=\identity_x\otimes\sigma_1$, where $\identity_x=\sum_x\ket{x}\bra{x},\ \identity_s=\sum_{s=L,R}\ket{s}\bra{s}$, and $\mathcal{K}$ is the complex conjugation operator. Thus, $U_1$ and $U_2$ are classified into BDI class \cite{asboth2012symmetries,asboth2013bulk,obuse2015unveiling,gerasimenko2016attractor,
mochizuki2016explicit}. It is known that, if we consider the system depicted in Fig. \ref{fig:theta}, there exist topologically protected edge states with $\varepsilon=0$ and $\varepsilon=\pi$ near $x=\pm m$ where $\theta_i(x)$ changes its sign. We describe edge states in single and two-step quantum walks as
\begin{equation}
\ket{\Phi_{\varepsilon,\eta}}=
\sum_{x,s}\Phi_{x,s}^{(\varepsilon,\eta)}
\ket{x}\otimes\ket{s},
\label{eq:edgestates_notation}
\end{equation}
where the labels $\varepsilon=0,\pi$ and $\eta=\pm$ represent the quasienergy and chirality of edge states, respectively. Thereby, the edge states $\ket{\Phi_{\varepsilon,\eta}}$ satisfy
\begin{equation}
U_{1/2}\ket{\Phi_{\varepsilon,\eta}}=
e^{-i\varepsilon}\ket{\Phi_{\varepsilon,\eta}},\ \ 
\Gamma\ket{\Phi_{\varepsilon,\eta}}=
\eta\ket{\Phi_{\varepsilon,\eta}}.
\label{eq:varepsilon_eta}
\end{equation}
Since the chiral symmetry operator is $\Gamma=\identity_x\otimes\sigma_1$, wave function amplitudes of edge states satisfy the relation
\begin{equation}
\Phi_{x,L}^{(\varepsilon,\eta)}
=\eta\Phi_{x,R}^{(\varepsilon,\eta)}.
\label{eq:LR_edgestates}
\end{equation}
By taking Eq. (\ref{eq:LR_edgestates}) into account, we use a simplified notation
\begin{equation}
 |\Phi_x^{(\varepsilon)}|\equiv|\Phi_{x,L}^{(\varepsilon,\eta)}|=|\Phi_{x,R}^{(\varepsilon,\eta)}|,
 \label{eq:Phi}
\end{equation}
when we focus on absolute values of wavefunction amplitudes and $\varepsilon$.

\section{Nonlinear quantum walks and stability of edge states: previous work}
\label{sec:previous}
It has been proposed to introduce nonlinear effects in quantum walks, using feed-forward control in optical systems. In the proposed feed-forward scheme \cite{shikano2014discrete,lee2015quantum}, the intensity of light is measured during the propagation and the conditions of optical elements corresponding to coin operators through which light passes after the measurement are changed depending on results of the measurement. In Ref. \cite{gerasimenko2016attractor}, a nonlinear coin operator
\begin{equation}
C(\kappa \Theta)=\sum_x \ket{x}\bra{x} \otimes
{\mathcal C}[\kappa \Theta(x,t)]
\label{eq:nonlinear_coin_operator}
\end{equation}
was introduced, where $\Theta(x,t)$ corresponding to the rotation angle of
the coin operator is defined as
\begin{equation}
\Theta(x,t)= 
|\psi_{x,L}(t)|^2
-|\psi_{x,R}(t)|^2.
\label{eq:Q_xt}
\end{equation}
Since $\Theta(x,t)$ depends on wave function amplitudes at every
time step $t$, the coin operator $C(\kappa \Theta)$ introduces nonlinear effects. The parameter $\kappa$ denotes the strength of nonlinearity. For convenience, we fix the sign of $\kappa$ as $\kappa\geq0$ in the following. From Eqs. (\ref{eq:LR_edgestates}), (\ref{eq:nonlinear_coin_operator}), and (\ref{eq:Q_xt}), we can understand that the nonlinear coin operator has no effect on edge states, i.e. $C(\kappa \Theta)\ket{\Phi_{\varepsilon,\eta}}=C(0)\ket{\Phi_{\varepsilon,\eta}}=\ket{\Phi_{\varepsilon,\eta}}$, because of $|\Phi_{x,L}^{(\varepsilon,\eta)}|^2-|\Phi_{x,R}^{(\varepsilon,\eta)}|^2=0$. Therefore, edge states are stationary states in the nonlinear quantum walks we consider below. The time evolution of a single-step nonlinear quantum walk is described by $U_1$ and $C(\kappa \Theta)$,
\begin{equation}
\ket{\psi(t+1)}=U_1C(\kappa \Theta)\ket{\psi(t)}.
\label{eq:time_evolution_1step}
\end{equation}
Since $\Theta$ introduces nonlinear effects,
Eq. (\ref{eq:time_evolution_1step}) is a nonlinear equation. As
mentioned above, edge states $\ket{\Phi_{\varepsilon,\eta}}$ are
stationary states in this dynamics. Taking the continuum limit in $x$
and $t$, Eq. (\ref{eq:time_evolution_1step}) is transformed into a nonlinear Dirac equation
\begin{equation}
i\frac{\partial}{\partial t}\ket{\psi_x(t)}=
-i\sigma_3\frac{\partial}{\partial x}\ket{\psi_x(t)}
+[\theta_0(x)+\kappa \Theta(x,t)]
\sigma_2\ket{\psi_x(t)},
\label{eq:continuum_limit}
\end{equation}
where $\ket{\psi_x(t)}=[\psi_{x,L}(t),\psi_{x,R}(t)]^{\rm T}$ and only
the first order in $\theta_0(x)$ and $\kappa \Theta(x,t)$ is taken into
account. Using Eq. (\ref{eq:continuum_limit}), the stability of edge states
is studied in Ref.  \cite{gerasimenko2016attractor}.\\\indent
Suppose that $\ket{\psi(t)}$ consists of the stationary edge state with quasienergy $\varepsilon$ and chirality $\eta$, $\ket{\Phi_{\varepsilon,\eta}}$, and an infinitesimally weak fluctuating state around the edge state $\ket{\delta\psi(t)}$,
\begin{equation}
\ket{\psi(t)}
=\ket{\Phi_{\varepsilon,\eta}}
+\ket{\delta\psi(t)},\ 
\ket{\delta\psi(t)}=\sum_{x,s}
\delta\psi_{x,s}(t)\ket{x}\otimes\ket{s},
\label{eq:seperate_1step}
\end{equation}
where $|\delta\psi_{x,s}(t)/\Phi_{x,s}^{(\varepsilon,\eta)}| \ll 1$ is assumed. Substituting Eq.\ (\ref{eq:seperate_1step}) into Eq.\ (\ref{eq:continuum_limit}) and expanding the equation up to the first order in $\delta\psi_{x,s}(t)$, the time-evolution equation for $\ket{\delta\psi(t)}$
\begin{equation}
i\frac{\partial}{\partial t}\ket{\delta\psi_x(t)}=\Omega\ket{\delta\psi_x(t)},\ 
\Omega=-i\sigma_3\frac{\partial}{\partial x}+\theta_0\sigma_2
-2i\kappa|\Phi|^2[\eta\identity_s-\sigma_1]
\label{eq:continuum_limit_distribution}
\end{equation}
is obtained, where
$\ket{\delta\psi_x(t)}=[\delta\psi_{x,L}(t),\delta\psi_{x,R}(t)]^{\rm T}$. In
Eq. (\ref{eq:continuum_limit_distribution}), for simplicity, $x$
dependence of $\theta_0(x)$ is ignored. In addition, $x$ and
$\varepsilon$ dependences of $|\Phi_x^{(\varepsilon)}|$ are also
ignored, and we write it as $|\Phi|$. From
Eq. (\ref{eq:continuum_limit_distribution}), we can understand that the
time evolution of a plane wave state with wave number $q$,
$\ket{\delta\psi_x(t)}=e^{i(qx-\omega t)}\ket{\delta\psi_0(0)}$, is determined by the complex frequency
\begin{equation}
\omega=-2i\kappa\eta|\Phi|^2\pm\sqrt{q^2+\theta_0^2-4\kappa^2|\Phi|^4},
\label{eq:complex-frequency}
\end{equation}
which is the eigenvalue of $\Omega$ (see \ref{sec:continuum_limit} for details). On one hand, if all of Im($\omega$) are negative, $\ket{\delta\psi_x(t)}$ decays with time steps. Then, $\ket{\Phi_{\varepsilon,\eta}}$ remains as a stable state. Therefore, Eq. (\ref{eq:complex-frequency}) means that an edge state with chirality $\eta=+$ is always stable when $\kappa>0$. On the other hand, if there is $\omega$ whose imaginary part is positive, then, $\ket{\delta\psi_x(t)}$ grows with time steps and overhelms the edge state $\ket{\Phi_{\varepsilon,\eta}}$. In this case, $\ket{\Phi_{\varepsilon,\eta}}$ is unstable. So, when $\kappa>0$, $\ket{\Phi_{\varepsilon,-}}$ are always repellers. From the analysis done in Ref. \cite{gerasimenko2016attractor}, it is concluded that the strength of nonlinearity $\kappa\ (>0)$ is irrelevant to determining whether an edge state becomes an attractor or a repeller. \\\indent
It is worth noting that, $\Omega$ in Eq. (\ref{eq:continuum_limit_distribution}) can be seen as a non-Hermitian Hamiltonian and possesses so-called $\mathcal{PT}$ symmetry. However, $\mathcal{PT}$ symmetry does not have any effect on the stability of edge states, while $\mathcal{PT}$ symmetry breaking occurs as the value of $\kappa$ is varied. See \ref{sec:continuum_limit} for details.

\section{Nonlinear quantum walks and stability of edge states: present work}
\label{sec:present}
In this section, we explore the stability of edge states in single and
two-step nonlinear quantum walks, without taking the continuum limit,
which differs from the analysis in Ref.\ \cite{gerasimenko2016attractor} explained in Sec. \ref{sec:previous}. We will demonstrate below that this scheme is essential to find additional bifurcations unique to Floquet systems.\\\indent
In both single and two-step nonlinear quantum walks, there are cases where soliton-like states appear when an initial state $\ket{\psi(0)}$ is localized at a single site, which we do not focus on in the present work. Since we empirically know that soliton-like states do not appear if $\ket{\psi(0)}$ is a Gaussian wave packet, we employ the following initial state 
\begin{equation}
\psi_{x,L}(0)=\psi_{x,R}(0)=N\exp(-x^2/2\Delta^2),
\label{eq:initial_state}
\end{equation}
where the standard deviation $\Delta$ is a parameter and $N$ is a real normalization constant. As wave function amplitudes of the initial state are all real, $\psi_{x,s}(t)$ are always real during the nonlinear time evolution.
\subsection{single-step nonlinear quantum walks}
\label{subsec:1step}
First, we consider the single-step nonlinear quantum walk explained in Sec. \ref{sec:previous}, the same model in Ref.  \cite{gerasimenko2016attractor}. Substituting Eq. (\ref{eq:seperate_1step}) into Eq. (\ref{eq:time_evolution_1step}) we obtain the time-evolution equation for $\ket{\delta\psi(t)}$,
\begin{equation}
\ket{\delta\psi(t+1)}=V_1^{(\varepsilon,\eta)}
\ket{\delta\psi(t)},\ V_1^{(\varepsilon,\eta)}=U_1 D_{\kappa,\eta}(|\Phi_x^{(\varepsilon)}|),
\label{eq:linearize_1step}
\end{equation}
up to the first order in $\delta\psi_{x,s}(t)$, where $D_{\kappa,\eta}(|\Phi_x^{(\varepsilon)}|)$ is
\begin{equation}
D_{\kappa,\eta}(|\Phi_x^{(\varepsilon)}|)=\sum_x\ket{x}\bra{x}
\otimes\mathcal{D}_{\kappa,\eta}(|\Phi_x^{(\varepsilon)}|),
\label{eq:D_definition}
\end{equation}
\vspace{-7mm}
\begin{equation}
\hspace{-17mm}\mathcal{D}_{\kappa,\eta}(|\Phi_x^{(\varepsilon)}|)=
\identity_s-2\kappa|\Phi_x^{(\varepsilon)}|^2
\left(\begin{array}{cc}
\eta&-1\\
-1&\eta
\end{array}\right)
=(1-2\kappa\eta|\Phi_x^{(\varepsilon)}|^2)\identity_s
+2\kappa|\Phi_x^{(\varepsilon)}|^2\sigma_1.
\label{eq:Dx_definition}
\end{equation}
We remark that, if $\kappa=0$, $D_{\kappa,\eta}(|\Phi_x^{(\varepsilon)}|)$ is an identity
operator and $V_1^{(\varepsilon,\eta)}=U_1$ is a unitary operator. Then,
the eigenvalues of $V_1^{(\varepsilon,\eta)},\ \lambda_1^{(\varepsilon,\eta)}$, are complex numbers on a unit circle in a complex plane. For $\kappa>0$, since $D_{\kappa,\eta}(|\Phi_x^{(\varepsilon)}|)$ and then $V_1^{(\varepsilon,\eta)}$ are nonunitary operators, $\lambda_1^{(\varepsilon,\eta)}$ are not on the unit circle in general. We note that $\lambda_1^{(\varepsilon,\eta)}$ appears as
quartets because of symmetries of the non-unitary time-evolution operator $V_1^{(\varepsilon,\eta)}$. Due to the bipartite structure of $V_1^{(\varepsilon,\eta)}$, $V_1^{(\varepsilon,\eta)}$ has sublattice symmetry ${\mathcal S} V_1^{(\varepsilon,\eta)} {\mathcal S}^{-1}=-V_1^{(\varepsilon,\eta)}$, where ${\mathcal S}$ is a unitary operator. In addition, by following the symmetry classification of non-Hermitian systems in Ref.\ \cite{kawabata2019symmetry}, the non-unitary
operator $V_1^{(\varepsilon,\eta)}$ possesses AZ$^\dagger$ particle-hole symmetry $[\tilde{\Xi} (V_1^{(\varepsilon,\eta)})^* \tilde{\Xi}^{-1}]=V_1^{(\varepsilon,\eta)}$, where $\tilde{\Xi}$ is a unitary operator. These two symmetries guarantee that eigenvalues appear as quartets, $\pm\lambda_1^{(\varepsilon,\eta)}$ and $\pm(\lambda_1^{(\varepsilon,\eta)})^*$.

In the linear stability analysis, eigenstates which have the largest value of $|\lambda_1^{(\varepsilon,\eta)}|$ dominate the dynamics of the nonunitary time evolution in Eq. (\ref{eq:linearize_1step}). On one hand, if max$(|\lambda_1^{(\varepsilon,\eta)}|)\leq1$, $|\delta\psi_{x,s}(t)|$ decays with time steps or remains small. Then, the edge state $\Ket{\Phi_{\varepsilon,\eta}}$ becomes a stable attractor. On the other hand, if max$(|\lambda_1^{(\varepsilon,\eta)}|)>1$, $|\delta\psi_{x,s}(t)|$ grows with time steps and $\ket{\Phi_{\varepsilon,\eta}}$ becomes an unstable repeller. Taking it into account that $U_1$ is a unitary operator
and $D_{\kappa,\eta}(|\Phi_x^{(\varepsilon)}|)$ is Hermitian, the upper bound of
$\max(|\lambda_1^{(\varepsilon,\eta)}|)$ is determined by
\begin{equation}
 \max(|\lambda_1^{(\varepsilon,\eta)}|) \le
\max_x(1,|1-4\kappa \eta |\Phi_x^{(\varepsilon)}|^2|),
 \label{eq:max_lambda1}
\end{equation}
where $1$ and $1-4\kappa\eta|\Phi_x^{(\varepsilon)}|^2$ are eigenvalues of $\mathcal{D}_{\kappa,\eta}(|\Phi_x^{(\varepsilon)}|)$ (see \ref{sec:bifurcation_points} for derivations). While Eq.\ (\ref{eq:max_lambda1}) gives only the upper bound of $\max(|\lambda_1^{(\varepsilon,\eta)}|)$, as we numerically demonstrate later, this inequality can correctly estimate the criteria
of the linear stability analysis, namely, $\max(|\lambda_1^{(\varepsilon,\eta)}|) \le 1$ or
$\max(|\lambda_1^{(\varepsilon,\eta)}|) > 1$, at least for the present
model. Thereby, here we summarize the results derived from Eq.\
(\ref{eq:max_lambda1}) under the facts that $\kappa>0$ and
$|\Phi_x^{(\varepsilon)}|>0$:
\begin{enumerate}
 \item In the case of $\eta=-$, since max$(|\lambda_1^{(\varepsilon,-)}|)\leq 1+4\kappa\max_x(|\Phi_x^{(\varepsilon)}|^2)>1$ is satisfied for any $\kappa$, $\ket{\Phi_{\varepsilon,-}}$ can always be unstable repellers.
\item  In the case of $\eta=+$, $\ket{\Phi_{\varepsilon,+}}$ are inevitably stable attractors when $\kappa$ is smaller than a threshold value $\kappa_c$, which guarantees
       max$(|\lambda_1^{(\varepsilon,+)}|)\leq1$. For $\kappa>\kappa_c$, however, an additional bifurcation can occur since max$(|\lambda_1^{(\varepsilon,+)}|)$ can be larger than one.
 \item The threshold value $\kappa_c$ is given by
 \begin{equation}
  \kappa_c = \frac{1}{2\max_x(|\Phi_x^{(\varepsilon)}|^2)},
\label{eq:kappa_c_1}
 \end{equation}
       which is obtained from the condition $1-4\kappa \eta\max_x(|\Phi_x^{(\varepsilon)}|^2)=-1$.
\end{enumerate}
The first result is consistent with the conclusion in Ref.\ \cite{gerasimenko2016attractor}.
However, the second and third results claim that $\ket{\Phi_{\varepsilon,+}}$ can be unstable for $\kappa>\kappa_c$ and an additional bifurcation occurs with increasing the strength of nonlinearity $\kappa$.

\begin{figure}[b]
\begin{center}
\includegraphics[scale=0.45]{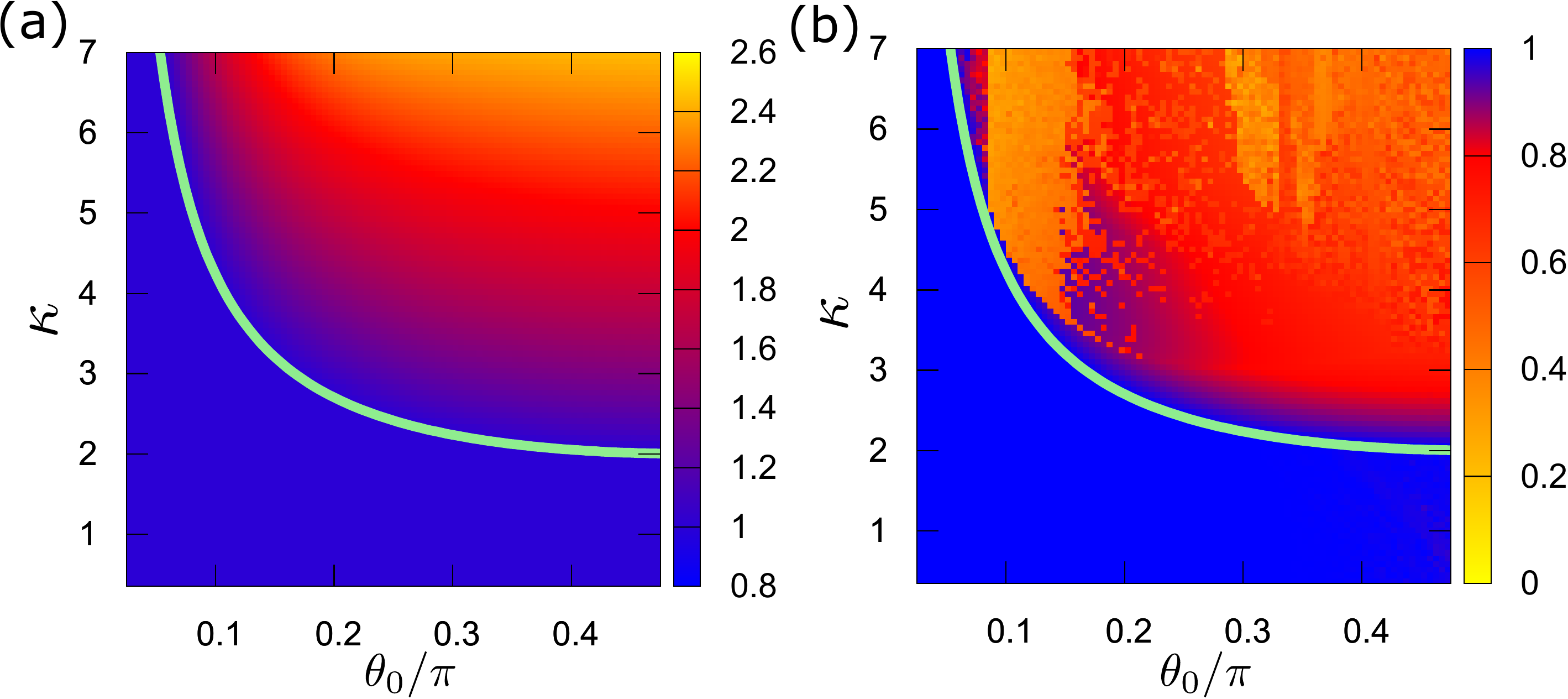}
\caption{$\theta_0$ and $\kappa$ dependences of
 (a)\ $\max(|\lambda_1^{(0,+)}|)$ and (b)\ $\max[F_{0,+}(T),F_{\pi,+}(T)]$ for $T=3\times10^6$.
 The threshold predicted from Eq.\ (\ref{eq:kappa_c_1}) is
 shown by the green solid curves. The initial state is given by Eq.\
 (\ref{eq:initial_state}) with $\Delta^2=30$ and $m=100$. }
\label{fig:phase_diagram_1step}
\end{center}
\end{figure}

In order to confirm the validity of Eq.\
(\ref{eq:kappa_c_1}) by numerically calculating eigenvalues of $V_1^{(\varepsilon,\eta)}$,
we derive the analytical solution of
$\ket{\Phi_{\varepsilon,\eta}}$. For simplicity, we consider the case
in which $\theta_0>0$ is satisfied. Under an assumption that left and right boundaries in Fig. \ref{fig:theta} locate far away and the overlap of edge states on each boundary is negligible, $\ket{\Phi_{\varepsilon,\eta}}$ becomes
\begin{equation}
\Phi^{(\varepsilon,\eta)}_{x,L}
=\eta\Phi^{(\varepsilon,\eta)}_{x,R}=
N_1(-1)^{\frac{\varepsilon}{\pi}x}
e^{-\gamma|x-\eta m^\prime|},
\label{eq:edgestates_1step}
\end{equation}
where an inverse of a localization length $\gamma$ and a normalization constant $N_1$ are
\begin{equation}
\gamma=\log(\frac{1+\sin\theta_0}{\cos\theta_0}),\ \ 
N_1=\sqrt{\frac{1-e^{-2\gamma}}{4}},
\label{eq:gamma_N_1step}
\end{equation}
and $m^\prime=m$ for $|x|\leq m$ and $m^\prime=m+1$ for $|x|\geq  m+1$. The derivation of Eqs. (\ref{eq:edgestates_1step}) and (\ref{eq:gamma_N_1step}) is given in \ref{sec:edgestates_1step}. Edge states with $\eta=+,\ \ket{\Phi_{\varepsilon,+}}$, are localized at the right boundary, and $\ket{\Phi_{\varepsilon,-}}$ are localized at the left boundary. From Eqs.\ (\ref{eq:kappa_c_1}) and (\ref{eq:edgestates_1step}), we obtain the expected threshold value $\kappa_c$. Alternating sign changes on the position space, $(-1)^{\frac{\varepsilon}{\pi}x}$, do not influence $D_{\kappa,\eta}(|\Phi_x^{(\varepsilon)}|)$ and $V_1^{(\varepsilon,\eta)}$, which can be understood from Eq. (\ref{eq:Dx_definition}).
\begin{figure}[tb]
\begin{center}
\includegraphics[width=10.5cm]{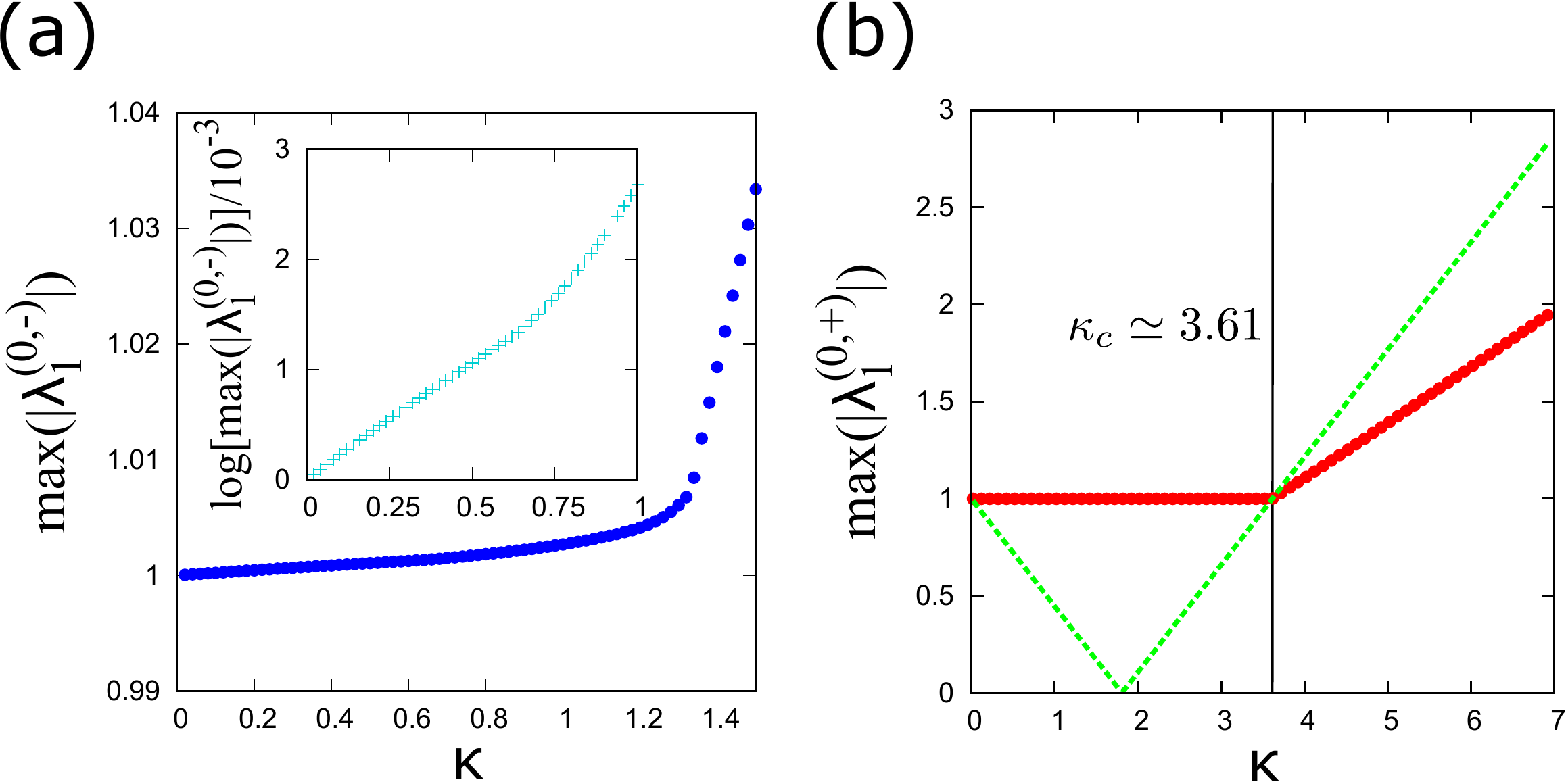}
\caption{$\kappa$ dependence of $\max(|\lambda_1^{(0,\eta)}|)$, with
 $\theta_0=\pi/8$ and $m=150$. (a) In the case of $\eta=-$. The
 inset shows the logarithm of $\max(|\lambda_1^{(0,-)}|)$. (b)\ In the case of $\eta=+$. The green dashed line represents $|1-4\kappa\max_x(|\Phi_x^{(\varepsilon)}|^2)|$ (see \ref{sec:bifurcation_points}). }
\label{fig:max_1step}
\end{center}
\end{figure}
Thus, in the single-step nonlinear quantum walk,
$V_1^{(0,\eta)}=V_1^{(\pi,\eta)}$ and
$\lambda_1^{(0,\eta)}=\lambda_1^{(\pi,\eta)}$ hold, since
$\ket{\Phi_{0,\eta}}$ and $\ket{\Phi_{\pi,\eta}}$ have the same
localization length $\gamma^{-1}$.  Therefore, we focus only on
$V_1^{(0,\eta)}$ and $\lambda_1^{(0,\eta)}$ hereafter. Because of this equivalence, however, there are always two stable states $\ket{\Phi_{0,\eta}}$ and $\ket{\Phi_{\pi,\eta}}$ when $\max(|\lambda_1^{(0,\eta)}|)=\max(|\lambda_1^{(\pi,\eta)}|)\leq1$, and we cannot predict which state is realized after long time evolution from the linear stability analysis we shall explain. Substituting Eq. (\ref{eq:edgestates_1step}) into Eqs. (\ref{eq:linearize_1step})-(\ref{eq:Dx_definition}), we
calculate $\lambda_1^{(\varepsilon,\eta)}$ by numerical
diagonalizations. Figure \ref{fig:max_1step} shows $\kappa$ dependence
of max$(|\lambda_1^{(0,\eta)}|)$. In the case of $\eta=-$,
max$(|\lambda_1^{(0,-)}|)>1$ is satisfied for all $\kappa>0$ as shown in Fig. \ref{fig:max_1step} (a). Therefore, $\ket{\Phi_{\varepsilon,-}}$ are always unstable repellers. In the case of $\eta=+$, max$(|\lambda_1^{(0,+)}|)$ remains to be $1$ when $\kappa$ is smaller
than a threshold value $\kappa_c \approx 3.61$ (when $\theta_0=\pi/8$), as shown in Fig. \ref{fig:max_1step} (b). In this case, $\ket{\Phi_{\varepsilon,+}}$ are stable attractors. However, for $\kappa>\kappa_c$, max$(|\lambda_1^{(0,+)}|)$ becomes larger than one, which means that $\ket{\Phi_{\varepsilon,+}}$ also become unstable. These numerical results completely agree with the theoretical predictions from Eq.\ (\ref{eq:max_lambda1}). In Fig. \ref{fig:max_1step} (b), the absolute value of
$1-4\kappa\max_x(|\Phi_x^{(\varepsilon)}|^2)$, one of the eigenvalues of $D_{\kappa,+}(|\Phi_x^{(\varepsilon)}|)$, is also plotted to support this conclusion. 
\begin{figure}[tb]
\begin{center}
\includegraphics[width=15.5cm]{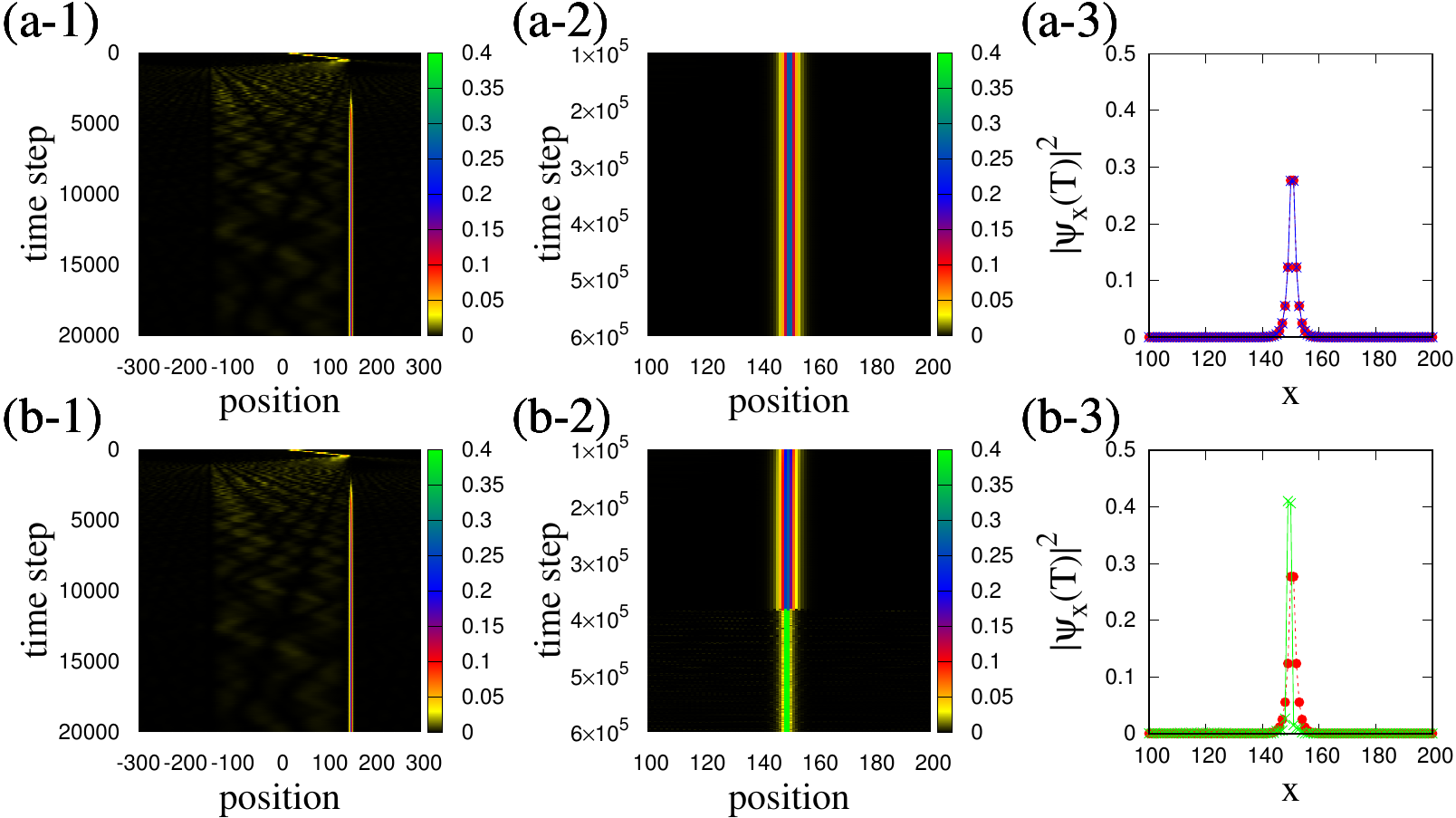}
\caption{Time evolution of the probability distribution (normalized intensity) $|\psi_x(t)|^2=|\psi_{x,L}(t)|^2+|\psi_{x,R}(t)|^2$ of the single-step nonlinear quantum walk with $\theta_0=\pi/8$ and $\Delta^2=50$. Two boundaries locate at $x=\pm 150$. The nonlinear parameter $\kappa$ is (a) below the threshold value $\kappa=3.6<\kappa_c$ and (b) above the threshold value $\kappa=3.9>\kappa_c$. The left and center columns show the time evolution in shorter and longer time scales, respectively. The right column shows probability distributions at the final time step $T=6\times10^5$, crosses and solid lines whose colors are blue and green, and the edge state $|\Phi_{0,+}\rangle$, red circles and dashed lines.}
 \label{fig:time-evolution_1step}
\end{center}
\end{figure}\\\indent
Here, we verify the additional bifurcation predicted from the stability analysis by numerically simulating time evolution of the nonlinear quantum walk in Eq.\ (\ref{eq:time_evolution_1step}) with the initial state in Eq.\ (\ref{eq:initial_state}). To begin with, we employ the same parameters as shown in Fig.\ \ref{fig:max_1step} where the stability analysis predicts the additional
bifurcation at a threshold value $\kappa_c\approx3.61$, from Eq. (\ref{eq:kappa_c_1}). Below the threshold, the probability distribution (normalized intensity) is accumulated around a boundary near $x=150$, forming a stationary state as shown in Fig. \ref{fig:time-evolution_1step} (a-1) and (a-2). Figure \ref{fig:time-evolution_1step} (a-3) compares the probability distribution of the stationary state at $T=6\times 10^5$ and that of an edge state $|\Phi_{0,+}\rangle$ of the quantum walk without nonlinearity. Since both states are almost identical, we can see that the edge state $|\Phi_{0,+}\rangle$ is a stable attractor for $\kappa<\kappa_c$. Above the threshold, the dynamics is similar to that with $\kappa<\kappa_c$ up to a certain time step; i.e. the probability distribution is accumulated around the boundary as shown in Fig. \ref{fig:time-evolution_1step} (b-1). Figure \ref{fig:time-evolution_1step} (b-2) clarifies that the shape of the probability distribution near the boundary abruptly shrinks around $3.8\times10^5$ time steps. Remarkably, the state after the abrupt shrink is not stationary as the state slightly fluctuates. From Fig.\ \ref{fig:time-evolution_1step} (b-3), we can see that the edge state $|\Phi_{0,+}\rangle$ is unstable for $\kappa>\kappa_c$, since the final state localizes stronger than the edge state $|\Phi_{0,+}\rangle$. As the probability distribution of the other edge state $|\Phi_{\pi,+}\rangle$ localized near $x=150$ is the same as that of $|\Phi_{0,+}\rangle$, we can also see that $|\Phi_{\pi,+}\rangle$ is unstable form Fig.\ \ref{fig:time-evolution_1step} (b-3).\\\indent
In order to quantitatively study the stability of edge states, we calculate a fidelity which is defined as
\begin{equation}
F_{\varepsilon,\eta}(t)
=|\naiseki{\Phi_{\varepsilon,\eta}}{\psi(t)}|.
\label{eq:fidelity_1step}
\end{equation}
If the edge state $|\Phi_{\varepsilon,\eta}\rangle$ is stable, the
fidelity is close to one after many time steps.
\begin{figure}[tb]
\begin{center}
\includegraphics[width=10.5cm]{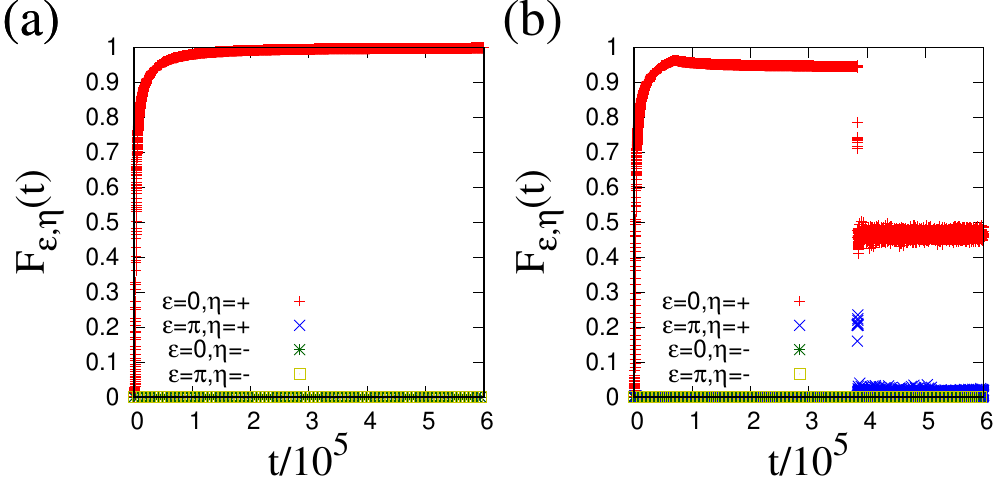}
 \caption{Time-step dependence of the fidelity $F_{\varepsilon,\eta}(t)$ for edge states
 $\ket{\Phi_{0,+}}$, $\ket{\Phi_{\pi,+}}$, $\ket{\Phi_{0,-}}$, and $\ket{\Phi_{\pi,-}}$. (a) $\kappa=3.6<\kappa_c$ and  (b) $\kappa=3.9>\kappa_c$, with $\theta_0=\pi/8,\ \Delta^2=50$, and $m=150$.}
\label{fig:fidelity_1step}
\end{center}
\end{figure}
In the case of $\kappa=3.6<\kappa_c$, as shown in Fig.\ \ref{fig:fidelity_1step} (a), the fidelity for the edge state $|\Phi_{0,+}\rangle$, $F_{0,+}(t)$, remains to be almost one after long
time steps. In the case of $\kappa=3.9>\kappa_c$ as shown in Fig.\ \ref{fig:fidelity_1step} (b), $F_{0,+}(t)$ and $F_{\pi,+}(t)$ cannot reach one and abruptly decrease around $3.8\times10^5$ time steps, which is consistent with the observation in Fig.\ \ref{fig:time-evolution_1step} (b-2).
Fidelities for other edge states with $\eta=-$ are almost zero during the time evolution. These
observations clearly validate the prediction from the linear stability analysis. We note that the abrupt shrink of the probability distribution in Fig. \ref{fig:time-evolution_1step} (b-2) or the sharp drop of $F_{0,+}(t)$ in Fig. \ref{fig:fidelity_1step} (b) cannot be predicted
from the linear stability analysis. This is because the linear stability analysis cannot predict behaviours in the unstable region where fluctuations become dominant, since the analysis assumes weak fluctuations around the edge states, while it can predict stable-unstable transitions.
\begin{figure}[b]
\begin{center}
\includegraphics[scale=0.55]{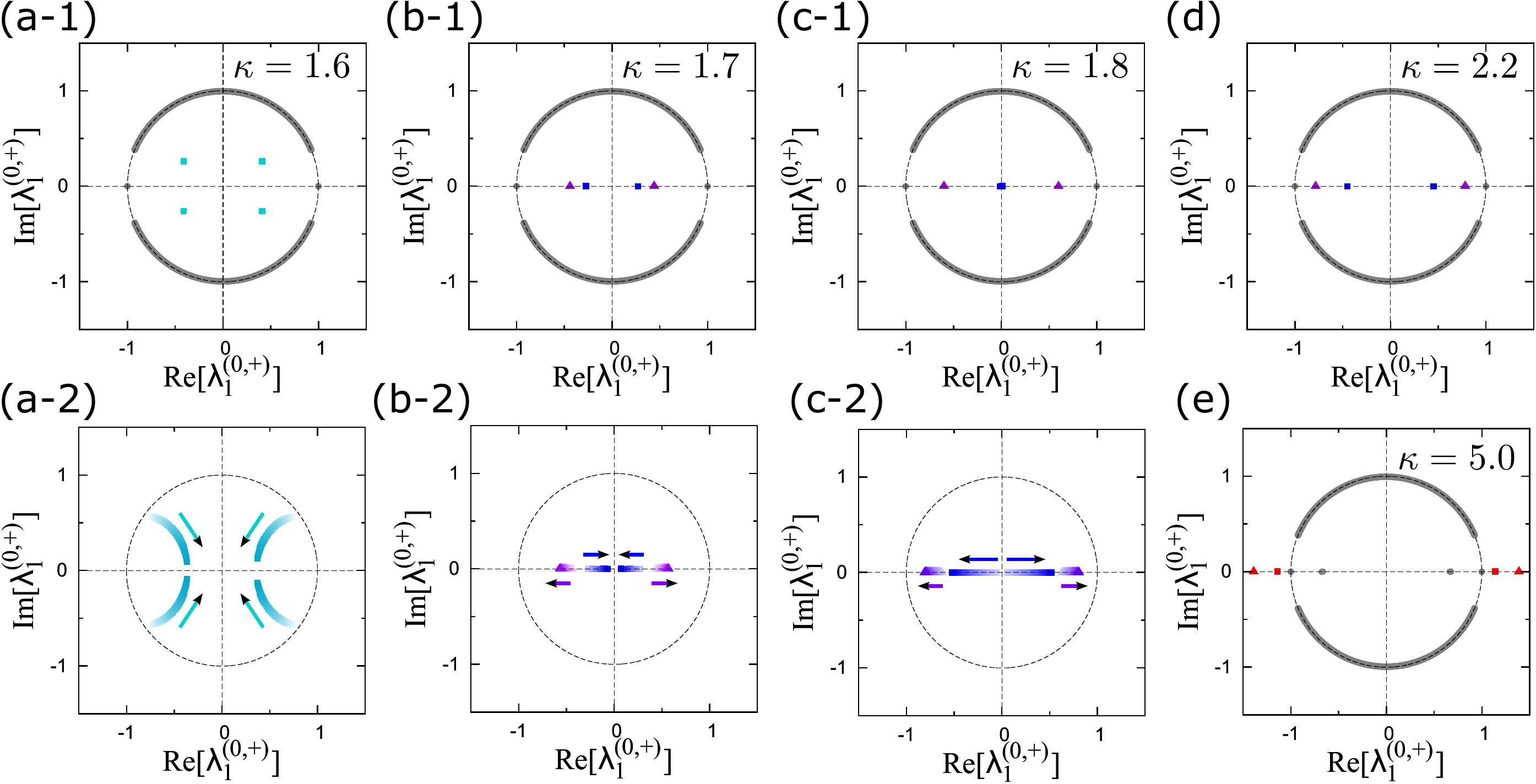}
\caption{Eigenvalues of $V_1^{(0,+)}$, $\lambda_1^{(0,+)}$, in the complex plane, with $\theta_0=\pi/8$ and $m=150$. Circles depicted in dashed lines are the unit circles. In (a-1)\ $\kappa=1.6$, (b-1)\ $\kappa=1.7$, (c-1)\ $\kappa=1.8$, (d)\ $\kappa=2.2$, and (e)\ $\kappa=5.0$, all eigenvalues are plotted. (a-2)\ $1\le\kappa\le1.68$, (b-2)\ $1.7\le\kappa\le1.77$, and (c-2)\ $1.85\le\kappa\le2.3$ show trajectories of particular four eigenvalues, which mainly contribute to the bifurcation of $\ket{\Phi_{\varepsilon,+}}$. The eigenvalues go toward the direction of the arrows as $\kappa$ is increased.}
\label{fig:eigenvalue_1step}
\end{center}
\end{figure}
\\\indent
For completeness, we explain how the additional bifurcation occurs, by showing all eigenvalues
of $V_1^{(0,+)},\ \lambda_1^{(0,+)}$, for various values of $\kappa$ in
the complex plane as shown in
Fig. \ref{fig:eigenvalue_1step}.
Here, the inside (outside) of the unit circle in the complex plane corresponds to a stable (unstable) region of the edge states.
First of all, we remark that almost all of $\lambda_1^{(0,+)}$ are identical or very close to eigenvalues of
$U_1$ which are on the unit circle due to unitarity of $U_1$, except a quartet of eigenvalues by following reasons. Firstly, the edge states of $U_1,\
\ket{\Phi_{\varepsilon,\eta}}$ are also eigenstates of
$V_1^{(\varepsilon,\eta)}$ with the same eigenvalue $e^{-i\varepsilon}$,
because $D_{\kappa,\eta}(|\Phi_x^{(\varepsilon)}|)\ket{\Phi_{\varepsilon,\eta}}=\ket{\Phi_{\varepsilon,\eta}}$
is satisfied, which can be understood from Eqs. (\ref{eq:D_definition})
and (\ref{eq:Dx_definition}). Secondly, ${\mathcal
D_{\kappa,\eta}(|\Phi_x^{(\varepsilon)}|)}\simeq\identity_s$ unless $x$ is near
the boundaries at $x=\pm m$, since $|\Phi_x^{(\varepsilon)}|^2$ takes
exponentially small values far away from the boundaries. Then, extended
bulk states of $U_1,\ \ket{\Psi_b}$, whose eigenvalues are $\mu_b$
satisfy $D_{\kappa,\eta}(|\Phi_x^{(\varepsilon)}|)\ket{\Psi_b}\simeq\ket{\Psi_b}$
and $V_1\ket{\Psi_b}\simeq\mu_b\ket{\Psi_b}$, as the bulk states have small
wave function amplitudes at $x=\pm m$. The only exception is
topologically trivial localized states, i.e. impurity states, which are
localized near boundaries but whose quasienergy is neither $0$ nor
$\pi$. These states take important roles for us to understand the additional
bifurcation as we explain in the next paragraph.\\\indent
We find that the absolute values of these eigenvalues are not equal to one and strongly depend on $\kappa$, the strength of nonlinearity. Increasing $\kappa$, the four eigenvalues, plotted as light blue squares, flow toward inside of the unit circle, and reach on the real axis [Fig. \ref{fig:eigenvalue_1step} (a)]. Further increasing $\kappa$, two eigenvalues (blue squares) flow toward the origin, while other two eigenvalues (purple triangles) move towards the opposite direction [Fig. \ref{fig:eigenvalue_1step} (b)]. After two eigenvalues (blue squares) collide and pass through each other at the origin [Fig. \ref{fig:eigenvalue_1step} (c)], four eigenvalues (purple triangles and blue squares) flow toward outside of the unit circle, being on the real axis [Fig. \ref{fig:eigenvalue_1step} (c-2),(d)]. Again increasing $\kappa$, the eigenvalues (plotted as red triangles and red squares) go out from the unit circle [Fig. \ref{fig:eigenvalue_1step} (e)], which makes $\ket{\Phi_{\varepsilon,+}}$ unstable. As we have explained, the motion of $\lambda_1^{(0,+)}$ in the complex plane changes $\ket{\Phi_{\varepsilon,+}}$ from stable to unstable
as $\kappa$ is increased. This additional bifurcation is related to that the stable region of $\lambda_1^{(\varepsilon,\eta)}$ is bounded,
which is one of peculiar features of Floquet systems (see also Sec. \ref{sec:discussion}). Furthermore, as mentioned in Sec. \ref{sec:previous}, the linear stability analysis in the continuum limit does not predict the additional bifurcation. These facts suggest that the bifurcation we have shown is unique to Floquet systems. \\\indent
Before closing this subsection, we extensively check the validity of the additional bifurcation.
Figures \ref{fig:phase_diagram_1step} (a) and (b) show $\max(|\lambda_1^{(0,+)}|)$ and $\max[F_{0,+}(T), F_{\pi,+}(T)]$ at $T=3\times 10^6$, respectively, for various values of $\theta_0$ and
$\kappa$. In Fig. \ref{fig:phase_diagram_1step} (b), as the linear stability analysis cannot predict which state is realized in the nonlinear dynamics, $\ket{\Phi_{0,+}}$ or $\ket{\Phi_{\pi,+}}$, we show $\max[F_{0,+}(T),F_{\pi,+}(T)]$. We also plot analytically predicted values of $\kappa_c$ in Eq.\ (\ref{eq:kappa_c_1}). We remark that the predicted values agree well with those obtained by numerical results. We also note that, as shown in Fig. \ref{fig:phase_diagram_1step}, $\kappa_c$ is a decreasing function of $\theta_0$ in $0<\theta_0<\pi/2$. This is because, when $\theta_0$ and the gap size for $\varepsilon=0,\pi$ are large, $|\Phi_x^{(\varepsilon)}|^2$ have a large value at localization positions $\pm m$ and $\pm(m+1)$, since the localization length $\gamma^{-1}$ is small (see \ref{sec:bifurcation_points} for details).

\subsection{two-step nonlinear quantum walks}
\label{subsec:2step}
Next, we consider a two-step nonlinear quantum walk whose time evolution is described as
\begin{equation}
\ket{\psi(t+1)}=U_{2b}C[-\kappa \tilde{\Theta}]
\ket{\tilde{\psi}(t)},\ \ket{\tilde{\psi}(t)}=U_{2a}C[\kappa \Theta]\ket{\psi(t)}.
\label{eq:time_evolution_2step}
\end{equation}
Here $\Theta(x,t)$ is defined from wave function amplitudes of $\ket{\psi(t)}$ as shown in Eq.\
(\ref{eq:Q_xt}). $\tilde{\Theta}(x,t)$ is also defined from $\ket{\tilde{\psi}(t)}$ in a simillar way:
\begin{equation}
\tilde{\Theta}(x,t)=|\tilde{\psi}_{x,L}(t)|^2
-|\tilde{\psi}_{x,R}(t)|^2.
\label{eq:theta_tilde}
\end{equation}
For the same reason with the single-step nonlinear quantum walk, edge states $\ket{\Phi_{\varepsilon,\eta}}$ are stationary states in this dynamics because of $\Theta(x,t)=\tilde{\Theta}(x,t)=0$ for $\ket{\psi(t)}=\ket{\Phi_{\varepsilon,\eta}}$. We emphasize that we cannot apply the analysis in terms of the effective Hamiltonian in Ref.\  \cite{gerasimenko2016attractor} to the two-step nonlinear quantum walk. This is because two nonlinear coin operators $C(\kappa \Theta)$ and $C(-\kappa\tilde{\Theta})$ in Eq.\ (\ref{eq:time_evolution_2step}) cancel out each other up to the first order in $\kappa$ by
taking the continuum limit in $x$ and $t$. Therefore, we have to directly analyze the time-evolution operator in order to check the stability of edge states in the two-step nonlinear quantum walk.

\begin{figure}[tb]
\begin{center}
\includegraphics[scale=0.9]{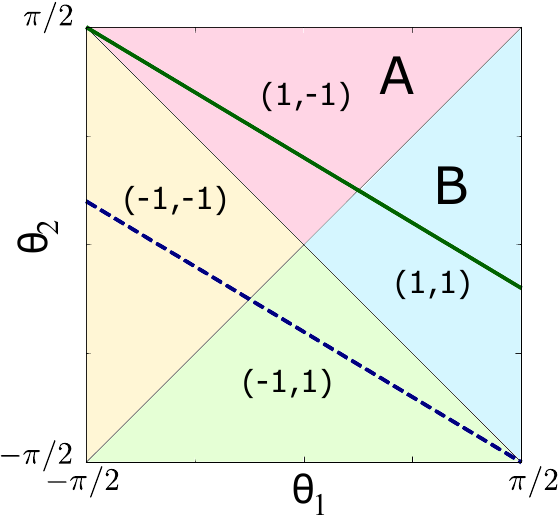}
\caption{$\theta_1$ and $\theta_2$ dependences of topological numbers
 ($\nu_0,\nu_\pi$). The green thick solid line, $\theta_2=-0.6\theta_1+\pi/5$, is used when the validity of the linear stability analysis is confirmed in Fig. \ref{fig:phase_diagram_2step}. Then, the coin parameters in the inner region in Fig.\ \ref{fig:theta} locate in the region A or B. The blue thick dashed line indicates $-\theta_1$ and $-\theta_2$ used for the outer regions in Fig.\ \ref{fig:theta}.}
\label{fig:theta12}
\end{center}
\end{figure}

Before analyzing the stability of edge states $\ket{\Phi_{\varepsilon,\eta}}$, we explain properties of
$\ket{\Phi_{\varepsilon,\eta}}$ for the two-step quantum walk without nonlinearlity in Eq.\ (\ref{eq:time_evolution_operator_2step}). We assume that two boundaries in Fig.\ref{fig:theta} are separated enough and edge states localized at different boundaries have no
overlap each other. See \ref{sec:edgestates_2step} for details. Note that, in the two-step quantum walk, edge states have finite wave function amplitudes only at even
sites or odd sites because of the decoupling between even and odd
sites. While we use $\ket{\Phi_{\varepsilon,\eta}}$ at even sites for the linear
stability analysis, our conclusion does not depend on the
even-odd parity. Figure \ref{fig:theta12} shows how the values of $(\nu_0,\nu_\pi)$ depend on the coin parameters $\theta_1$ and $\theta_2$, where $\nu_0$ and $\nu_\pi$ are topological numbers for zero-energy gap and $\pi$-energy gap, respectively. In the following, we consider that the coin parameters in the inner region shown in Fig.\ \ref{fig:theta} sweep the thick line (green) in regions A and B as shown in Fig.\ \ref{fig:theta12} for simplicity.
In both regions A and B, $\nu_0$ has the same value and the analytical form of $\ket{\Phi_{0,\eta}}$ is
\begin{equation}
\Phi_{x,L}^{(0,+)}=\left\{\begin{array}{ll}
N_2e^{-\gamma_0|x-m|}& (x\leq m)\\ 
N_2 p e^{-\gamma_0|x-(m+2)|}& (x> m)
\end{array}\right.
\label{eq:edgestates_2step_0_+}
\end{equation}
\begin{equation}
\Phi_{x,L}^{(0,-)}=\left\{\begin{array}{ll}
N_2 p e^{-\gamma_0|x-(-m-2)|}& (x<-m)\\
N_2 e^{-\gamma_0|x-(-m)|}& (x\geq-m)
\end{array}\right.
\label{eq:edgestates_2step_0_-}
\end{equation}
where $\gamma_0,\ p$, and $N_2$ are
\begin{equation}
\hspace{-15mm}\gamma_0 =\frac{1}{2}\log\left[
\frac{(1+\sin\theta_1)(1+\sin\theta_2)}{\cos\theta_1\cos\theta_2}\right],\ 
p =\frac{\cos\theta_2}{1+\sin\theta_2},\ \ N_2=\sqrt{\frac{1-e^{-4\gamma_0}}{2(1+p^2)}.}
\label{eq:gamma-zero_p_N}
\end{equation}
See \ref{sec:edgestates_2step} for derivations. Irrespective of the parameter regions, $\ket{\Phi_{0,+}}$ and $\ket{\Phi_{0,-}}$ are localized at the right and left boundaries, respectively. On the other hand, for $\varepsilon=\pi$, the value of $\nu_\pi$ and localization centers of $\ket{\Phi_{\pi,\eta}}$ depend on which parameter region $\theta_1$ and $\theta_2$ exist in. In the region A, $\ket{\Phi_{\pi,\eta}}$ is
\begin{equation}
\Phi_{x,L}^{(\pi,+)}=\left\{\begin{array}{ll}
N_2p(-1)^\frac{x}{2}e^{-\gamma_\pi^{\rm A}|x-(-m-2)|}&(x<-m)\\ 
N_2(-1)^\frac{x}{2}e^{-\gamma_\pi^{\rm A}|x-(-m)|}& (x\geq-m)
\end{array}\right.
\label{eq:edgestates_2step_pi_+_A}
\end{equation}
\begin{equation}
\Phi_{x,L}^{(\pi,-)}=\left\{\begin{array}{ll}
N_2(-1)^\frac{x}{2}e^{-\gamma_\pi^{\rm A}|x-m|}& (x\leq m)\\ 
N_2p(-1)^\frac{x}{2}e^{-\gamma_\pi^{\rm A}|x-(m+2)|}& (x>m)
\end{array}\right.
\label{eq:edgestates_2step_pi_-_A}
\end{equation}
where
\begin{equation}
\gamma_\pi^A=\frac{1}{2}\log\left[
\frac{(1-\sin\theta_1)(1+\sin\theta_2)}{\cos\theta_1\cos\theta_2}\right].  
\label{eq:gamma_pi_A}
\end{equation}
Equations (\ref{eq:edgestates_2step_pi_+_A}) and
(\ref{eq:edgestates_2step_pi_-_A}) mean that $\ket{\Phi_{\pi,+}}$ is localized at the left boundary and $\ket{\Phi_{\pi,-}}$ is vice versa, in the region A. In the region B,
\begin{equation}
\Phi_{x,L}^{(\pi,+)}=\left\{\begin{array}{ll}
N_2p(-1)^\frac{x}{2}e^{-\gamma_\pi^{\rm B}|x-m|}& (x\leq m)\\
N_2(-1)^\frac{x}{2}e^{-\gamma_\pi^{\rm B}|x-(m+2)|}& (x>m)
\end{array}\right.
\label{eq:edgestates_2step_pi_+_B}
\end{equation}
\begin{equation}
\Phi_{x,L}^{(\pi,-)}=\left\{\begin{array}{ll}
N_2(-1)^\frac{x}{2}e^{-\gamma_\pi^{\rm B}|x-(-m-2)|}& (x<-m)\\ 
N_2p(-1)^\frac{x}{2}e^{-\gamma_\pi^{\rm B}|x-(-m)|}& (x\geq-m)
\end{array}\right.
\label{eq:edgestates_2step_pi_-_B}
\end{equation}
where
\begin{equation}
\gamma_\pi^B=\frac{1}{2}\log\left[
\frac{(1+\sin\theta_1)(1-\sin\theta_2)}{\cos\theta_1\cos\theta_2}\right]. 
\label{eq:gamma_pi_B}
\end{equation}
Localization centers of $\ket{\Phi_{\pi,+}}$ and
$\ket{\Phi_{\pi,-}}$ in the region B are the opposite of
those in the region A. Note that we can obtain $\ket{\Phi_{\varepsilon,\eta}}$ in odd sites by changing $m$, $m+2$, and $p$ to $m-1$, $m+1$, and $p^{-1}$, respectively, in Eqs. (\ref{eq:edgestates_2step_0_+})-(\ref{eq:edgestates_2step_pi_-_B}). In regions A and B, the normalization constant $N_2$ for $\ket{\Phi_{\pi,\eta}}$ is obtained by substituting $\gamma_\pi^A$ and $\gamma_\pi^B$ into $\gamma_0$ in Eq. (\ref{eq:gamma-zero_p_N}), respectively.

Now, we consider the linear stability analysis for edge states of the two-step
nonlinear quantum walk. Using the analytical form of $\ket{\Phi_{\varepsilon,\eta}}$,
Eqs. (\ref{eq:edgestates_2step_0_+})-(\ref{eq:edgestates_2step_pi_-_B}),
we linearize the time-evolution equation
(\ref{eq:time_evolution_2step}). In the same way with the
single-step nonlinear quantum walk, we assume the infinitesimally weak
fluctuating state in Eq.\ (\ref{eq:seperate_1step})
and substitute it into Eqs. (\ref{eq:time_evolution_2step}).
Ignoring higher order terms in $\delta\psi_{x,s}(t)$, we can obtain the time-evolution equation of $\ket{\delta\psi(t)}$,
\begin{equation}
\hspace{-15mm}\ket{\delta\psi(t+1)}=V_2^{(\varepsilon,\eta)}\ket{\delta\psi(t)},\ 
V_2^{(\varepsilon,\eta)}=
U_{2b}D_{-\kappa,\tilde{\eta}}(|\tilde{\Phi}_x^{(\varepsilon)}|)
U_{2a}D_{\kappa,\eta}(|\Phi_x^{(\varepsilon)}|),
\label{eq:linearize_2step}
\end{equation}
where $|\tilde{\Phi}_x^{(\varepsilon)}|$ denotes absolute values of
wave function amplitudes, in the same way with Eq.\ (\ref{eq:Phi}). Edge states $\ket{\tilde{\Phi}_{\varepsilon,\tilde{\eta}}}$ are defined by 
\begin{equation}
\ket{\tilde{\Phi}_{\varepsilon,\tilde{\eta}}}
 =U_{2a}\ket{\Phi_{\varepsilon,\eta}}.
\end{equation}
As explained in \ref{sec:chirality_2step}, $\ket{\tilde{\Phi}_{\varepsilon,\tilde{\eta}}}$ is also the eigenstate of the same chiral symmetry operator $\Gamma$ and the following
relation is satisfied:
\begin{equation}
\Gamma \ket{\tilde{\Phi}_{\varepsilon,\tilde{\eta}}}
=\tilde{\eta}\ket{\tilde{\Phi}_{\varepsilon,\tilde{\eta}}},\ \ 
\tilde{\eta}=e^{i\varepsilon}\eta.
\label{eq:chirality_Phi_tilde}
\end{equation}
Here, we emphasize that chiralities $\tilde{\eta}$ depend on
$\varepsilon$. In the same way as we explained in Sec.\ \ref{subsec:1step},
the largest value of $|\lambda_2^{(\varepsilon,\eta)}|$, where
$\lambda_2^{(\varepsilon,\eta)}$ are eigenvalues of
$V_2^{(\varepsilon,\eta)}$, dominates the time evolution in Eq.\
(\ref{eq:linearize_2step}). If $\max(|\lambda_2^{(\varepsilon,\eta)}|)>1\ (\leq1)$ is satisfied, where  $\ket{\delta\psi(t)}$ grows (does not grow) with time steps and $\ket{\Phi_{\varepsilon,\eta}}$ is unstable (stable). As derived in \ref{sec:bifurcation_points}, the upper bound of $\max(|\lambda_2^{(\varepsilon,\eta)}|)$ is determined by a product of the maximum value of eigenvalues of
$D_{\kappa,\eta}(|\Phi_x^{(\varepsilon)}|)$ and that of
$D_{-\kappa,\tilde{\eta}}(|\tilde{\Phi}_x^{(\varepsilon)}|)$. The eigenvalues of $\mathcal{D}_{\kappa,\eta}(|\Phi_x^{(\varepsilon)}|)$ and $\mathcal{D}_{-\kappa,\tilde{\eta}}(|\tilde{\Phi}_x^{(\varepsilon)}|)$ are 
\begin{equation}
\delta_1=1,\ \delta_2(x)=1-4\kappa\eta|\Phi_x^{(\varepsilon)}|^2,\ 
\tilde{\delta}_1=1,\ \tilde{\delta}_2(x)=1+4\kappa\tilde{\eta}|\tilde{\Phi}_x^{(\varepsilon)}|^2, 
\label{eq:eigenvalue_D_2step}
\end{equation}
respectively. Taking these eigenvalues into account, the upper bound of
$\max(|\lambda_2^{(\varepsilon,\eta)}|)$ is expressed as
\begin{equation}
\hspace{-1.3mm}
\max(|\lambda_2^{(\varepsilon,\eta)}|) \le
 \max_x\Big(\delta_1,|\delta_2(x)|\Big)
 \max_x\left(\tilde{\delta}_1,|\tilde{\delta}_2(x)|\right).
 \label{eq:max_lambda2}
\end{equation}
Contrary to the single-step nonlinear quantum walk in Sec. \ref{subsec:1step}, $\max(|\lambda_2^{(\varepsilon,\eta)}|)$ depends on the quasienergy $\varepsilon$, since the localization length of $\ket{\Phi_{0,\eta}}$ and $\ket{\Phi_{\pi,\eta}}$ are different.

\begin{figure}[bt]
\begin{center}
\includegraphics[width=15cm]{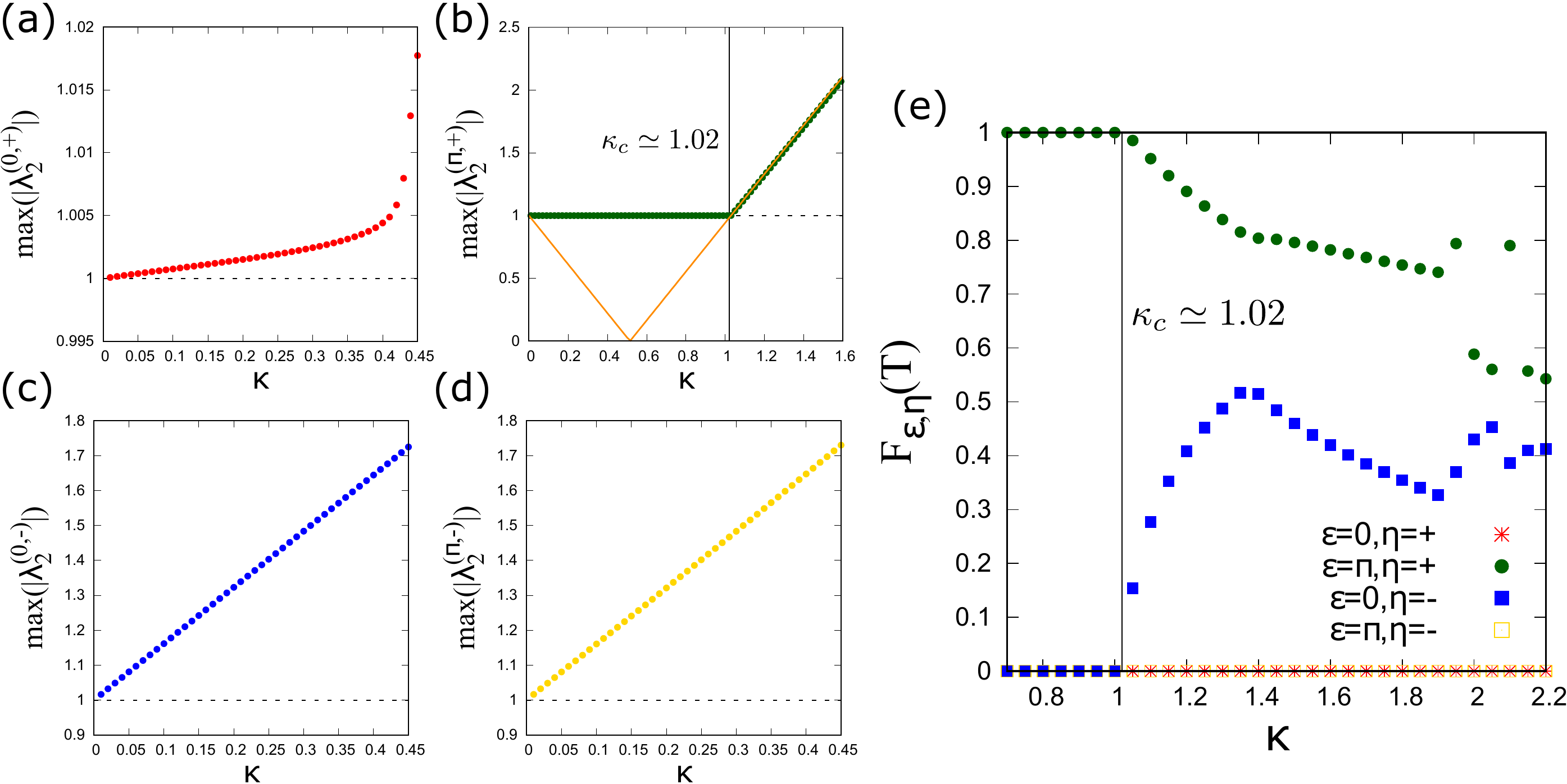}
\caption{(a)-(d)\ $\max(|\lambda_2^{(\varepsilon,\eta)}|)$ and (e)\ the
 fidelity at $T=10^7$, $F_{\varepsilon,\eta}(T)$, with
 $\theta_1=-\pi/4,\ \theta_2=2\pi/5$, $\Delta^2=50$, and $m=150$. The
 black dashed lines in (a)-(d) represent $\max(|\lambda_2^{(\varepsilon,\eta)}|)=1$, and the orange solid line in (d) represents $|1-4\kappa\eta\max_x(|\Phi_x^{(\varepsilon)}|^2)|$. Note that, $\ket{\psi(0)}$ and $\ket{\Phi_{\varepsilon,\eta}}$ have amplitudes only at even sites.}
\label{fig:max_fidelity_2step}
\end{center}
\end{figure}

\begin{figure}[b]
\begin{center}
\includegraphics[scale=0.55]{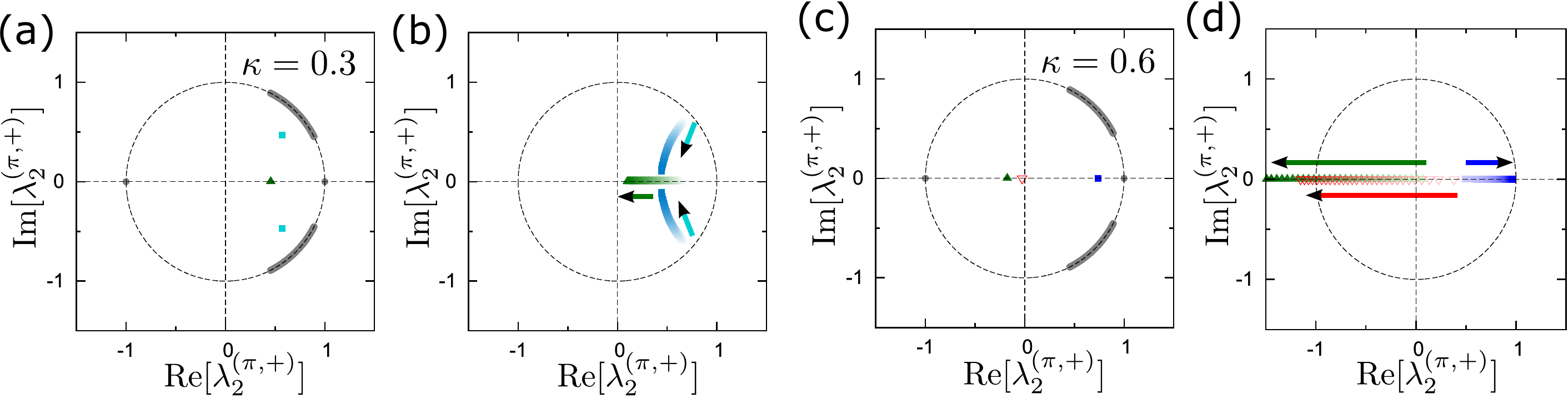}
\caption{Eigenvalues of $V_2^{(\pi,+)}$, $\lambda_2^{(\pi,+)}$, in the
 complex plane, with $\theta_1=-\pi/4,\ \theta_2=2\pi/5$, and
 $m=150$. The circles depicted in dashed lines are the unit
 circles. All eigenvalues are shown for (a)\ $\kappa=0.3$ and
 (c)\ $\kappa=0.6$. In (b)\ $0.2\le\kappa\le0.468$ and
 (d)\ $0.48\le\kappa\le1.288$, trajectories of three eigenvalues, which
 mainly contribute to the bifurcation of $\ket{\Phi_{\pi,+}}$, are
 plotted. The eigenvalues flow toward the direction of the arrows
 with increasing $\kappa$.}
\label{fig:eigenvalue_2step}
\end{center}
\end{figure}

\begin{figure}[b]
\begin{center}
\includegraphics[scale=0.45]{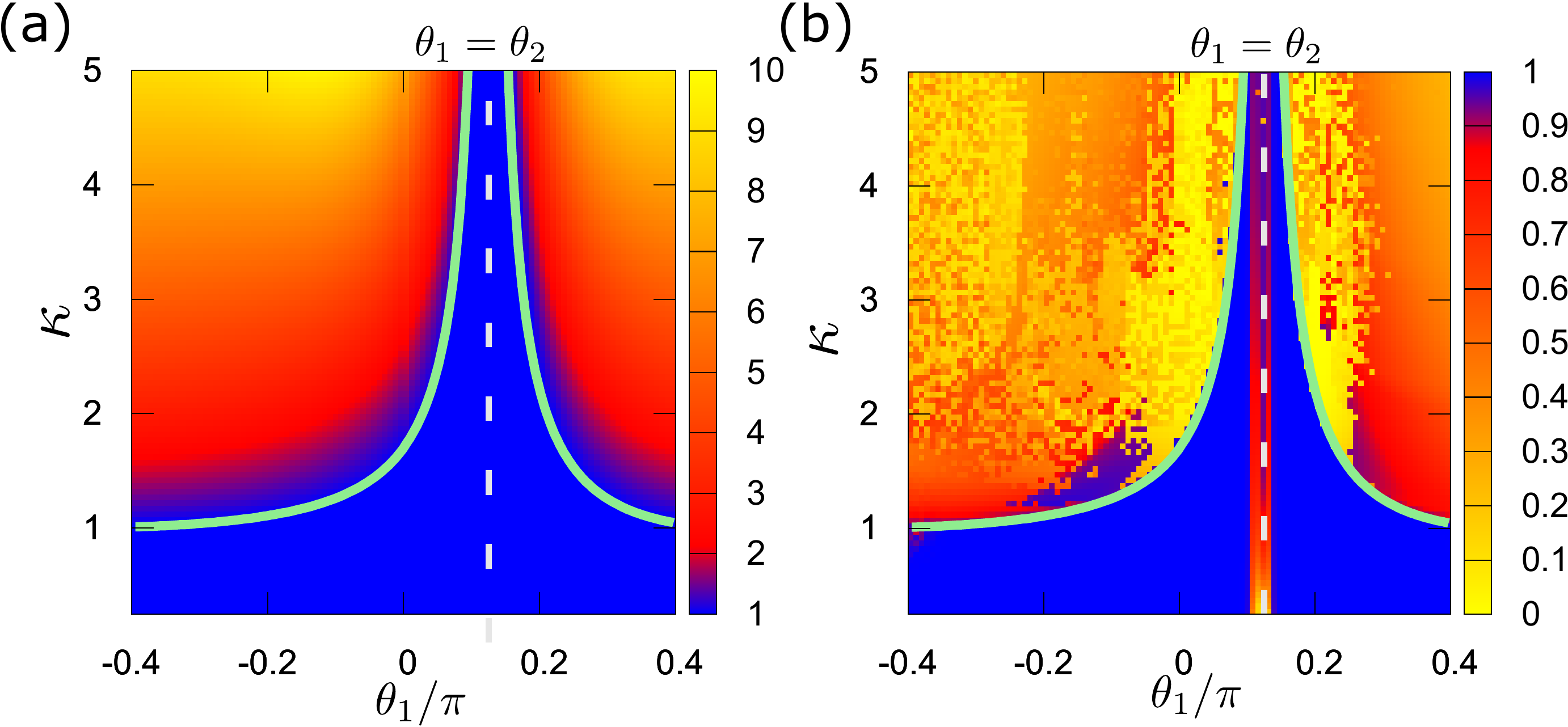}
\caption{$\theta_1$ and $\kappa$ dependences of
 (a)\ $\max(|\lambda_2^{(\pi,+)}|)$ and (b)\ $F_{\pi,+}(T)$ with $m=100,\ T=3\times10^6$, and $\Delta^2=70$. Note that, $\ket{\psi(0)}$ and $\ket{\Phi_{\pi,+}}$ have amplitudes only at odd sites. In both pictures (a) and (b), rotation angles $\theta_1$ and $\theta_2$ in the inner region (outer region) are scanned along the green solid (blue dashed) line in Fig. \ref{fig:theta12}, $\theta_2=-0.6\theta_1+\pi/5$, avoiding points on which $\theta_1=0,\ \theta_2=0$, or $\theta_1=\theta_2$. On the white dashed lines, $\theta_1=\theta_2=\pi/8$. The green lines represent $\kappa_c$ as a function of $\theta_1$, obtained from Eq. (\ref{eq:kappa_c_2}).}
\label{fig:phase_diagram_2step}
\end{center}
\end{figure}

Although Eq.\ (\ref{eq:max_lambda2}) gives only the upper bound of $\max(|\lambda_2^{(\varepsilon,\eta)}|)$, numerical calculations that we will show later guarantee that the
 inequality properly predicts the bifurcation points for the present model. Thereby, we again summarize the results obtained from Eq.\ (\ref{eq:max_lambda2}) here. 
\begin{enumerate}
 \item The edge states $\ket{\Phi_{0,\eta=\pm}}$ and
       $\ket{\Phi_{\pi,-}}$ can be unstable for any $\kappa>0$, since the right hand side of Eq. (\ref{eq:max_lambda2}) is always larger than one and $\max(|\lambda_2^{(\varepsilon,\eta)}|)>1$ can be satisfied.
\item  In the case of $\ket{\Phi_{\pi,\eta=+}}$, the edge state is inevitably stable as long as both $\max_x(|\delta_2(x)|)$ and $\max_x(|\tilde{\delta}_2(x)|)$ are smaller than or equal to one, corresponding to $\max(|\lambda_2^{(\pi,+)}|)\leq1$. This is satisfied when $\kappa$ is smaller than a threshold value $\kappa_c$. For $\kappa>\kappa_c$, $\max(|\lambda_2^{(\pi,+)}|)>1$ can be satisfied since $\max_x(|\delta_2(x)|)$ and/or $\max_x(|\tilde{\delta}_2(x)|)$ is larger than one. Therefore, a transition from a stable attractor to an unstable repeller occurs.
 \item The threshold is given by
\begin{equation}
 \kappa_c=\frac{1}{2\max_x\left(|\Phi_x^{(\pi)}|^2,|\tilde{\Phi}_x^{(\pi)}|^2\right)},
 \label{eq:kappa_c_2}
\end{equation}
which comes from the condition that one of $\max_x(|\delta_2(x)|)$ and $\max_x(|\tilde{\delta}_2(x)|)$ becomes one.
\end{enumerate}
Now, we numerically confirm the validity of the above results. To this end, we calculate eigenvalues of $V_2^{(\varepsilon,\eta)}$ by numerical diagonalizations after substituting the analytical solutions of edge states in Eqs.\ (\ref{eq:edgestates_2step_0_+})-(\ref{eq:gamma_pi_B}) into Eq.\ (\ref{eq:linearize_2step}). Figure \ref{fig:max_fidelity_2step} (a)-(d) show $\kappa$ dependence of the maximum value of $|\lambda_2^{(\varepsilon,\eta)}|$ for four kinds of edge states under specific conditions. We clearly observe that $\max(|\lambda_2^{(\varepsilon,\eta)}|)$ for edge states $\ket{\Phi_x^{0,+}}$, $\ket{\Phi_x^{0,-}}$, and $\ket{\Phi_x^{\pi,-}}$ become larger than one for $\kappa>0$. However, $\max(|\lambda_2^{(\pi,+)}|)$ for the edge state $\ket{\Phi_x^{\pi,+}}$ remains to be one as long as $\kappa<\kappa_c$, and starts to increase with increasing $\kappa$ further. When $\theta_1=-\pi/4$ and $\theta_2=2\pi/5$, the value of $\kappa_c$ is estimated as $\kappa_c\approx1.03$ from Eq.\ (\ref{eq:kappa_c_2}). These observations are consistent with the analytically derived predictions from Eq.\ (\ref{eq:max_lambda2}).
We verify the above result of the stability analysis by numerically
calculating the time evolution of the nonlinear quantum walk in Eq.\
(\ref{eq:time_evolution_2step}).
The fidelity $F_{\varepsilon,\eta}(T)$ at a time step $T=10^7$ is shown in Fig. \ref{fig:max_fidelity_2step} (e). As expected from $\max(|\lambda_2^{(\varepsilon,\eta)}|)$ in
Fig. \ref{fig:max_fidelity_2step} (a)-(d), $F_{\pi,+}(T)$ is almost
one (much smaller than one) for $\kappa\leq\kappa_c\
(\kappa>\kappa_c)$, while $F_{0,+}(T)$, $F_{0,-}(T)$, and
$F_{\pi,-}(T)$ are always much smaller than one.\\\indent
Here, we explain details for $\kappa$ dependences of $\lambda_2^{(\pi,+)}$, eigenvalues of
$V_2^{(\pi,+)}$, in the complex plane shown in Fig.\ \ref{fig:eigenvalue_2step}. Eigenvalues near the unit circle (gray circles) are inside or on the unit circle. Increasing $\kappa$, one real eigenvalue inside the unit circle (green triangle) flows toward the origin, while two complex eigenvalues (light blue squares) approach to the real axis\ [Fig. \ref{fig:eigenvalue_2step} (a) and (b)]. Further increasing $\kappa$, two eigenvalues (green triangle
and red empty triangle) flow toward the negative side and go
beyond the unit circle\ [Fig. \ref{fig:eigenvalue_2step} (c) and (d)].
This establishes $\kappa$ dependences of
$\max(|\lambda_2^{(\pi,+)}|)$ in Fig.\ \ref{fig:max_fidelity_2step}.
As the behavior of $\lambda_2^{(\pi,+)}$ in the complex plane is crucial, the bifurcation of $\ket{\Phi_{\pi,+}}$ in the nonlinear two-step quantum walk is one of the typical phenomena in Floquet systems.\\\indent
Finally, we present a comprehensive result to verify the additional bifurcation in the two-step nonlinear quantum walk. Figure \ref{fig:phase_diagram_2step} (a) and (b) show
$\max(|\lambda_2^{(\pi,+)}|)$ and $F_{\pi,+}(T)$ at $T=3\times10^6$, respectively, for various $\theta_1$, accordingly $\theta_2=-0.6\theta_1+\pi/5$, and $\kappa$. Comparing
Fig. \ref{fig:phase_diagram_2step} (a) with (b), it is obvious that the
stability analysis based on $\max(|\lambda_2^{(\pi,+)}|)$ gives the
correct prediction, excepting near $\theta_1=\theta_2=\pi/8$
where the topological number $\nu_\pi$ is not defined as shown in Fig.\
\ref{fig:theta12} and the edge states $\ket{\Phi_{\pi,\eta}}$ do not exist.
The reason is as follows. By putting $\theta_1=\theta_2$ into Eqs.\ (\ref{eq:gamma_pi_A}) and
 (\ref{eq:gamma_pi_B}), we obtain $\gamma_\pi^{A/B}=0$. Therefore,
 the localization length of the edge state diverges. Even near this point, $\ket{\Phi_{\pi,\eta}}$ have a huge localization length. Then, edge states localizing at left and right boundaries could largely overlap each other, and the assumption that $\ket{\Phi_{\pi,+}}$
 and $\ket{\Phi_{\pi,-}}$ are independent is not suitable. Therefore,
 we can shrink this exceptional region by making the system
 size larger. Note that, when $\ket{\Phi_{\pi,+}}$ is stable and the initial state $\ket{\psi(0)}$ in Eq. (\ref{eq:initial_state}) has amplitudes at both even and odd sites, $F_{\pi,+}(T)$ for large $T$ is almost $0.5$, since $\ket{\Phi_{\pi,+}}$ have amplitudes only at even or odd sites. Then, we need to replace $\ket{\Phi_{\pi,+}}$ with $\ket{\Phi_{\pi,+}}/\sqrt{2}$ in order to calculate $\lambda_2^{(\pi,+)}$, and this makes $\kappa_c$ become $2\kappa_c$.

\begin{figure}[t]
\begin{center}
\includegraphics[width=15cm]{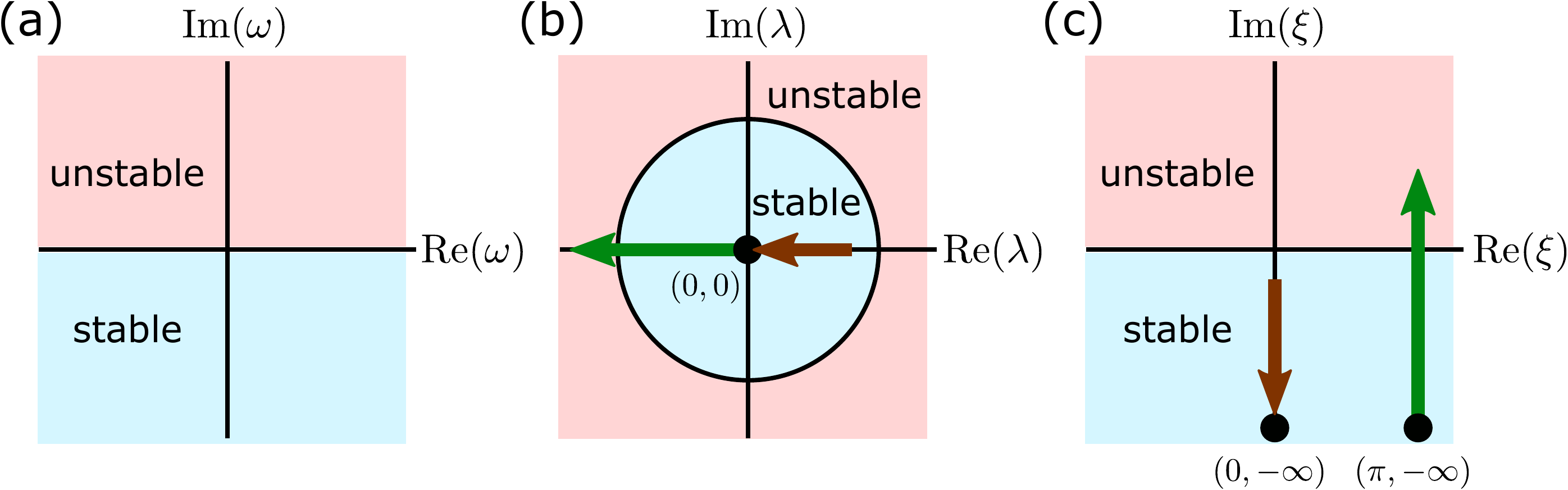}
\caption{Schematic pictures of stable and unstable regions in the complex plane, in the case of (a) usual systems (with no time-periodicity) and (b),(c) Floquet systems. In (b) and (c), brown and green arrows represent the motion of an eigenvalue of a time-evolution operator and quasienergy $\xi=i\log(\lambda)$, respectively. }
\label{fig:eigenvalue-motion}
\end{center}
\end{figure}

\section{Summary and discussion}
\label{sec:discussion}
In this work, we have shown that the stability of topologically protected edge states
of the single and two-step nonlinear quantum walks can be predicted by
the linear stability analysis, precisely taking the dynamical properties
of Floquet systems into consideration. Contrary to the previous work
\cite{gerasimenko2016attractor}, we have analytically found the additional
bifurcations depending on the strength of nonlinearity for both
nonlinear quantum walks by directly applying time-evolution operators of the quantum walks to the linear stability analysis. Then, we have confirmed the validity of the analysis by numerical calculations.\\\indent
Finally, we discuss the origin of the additional bifurcation from the
viewpoint of eigenvalues. In single and two-step nonlinear quantum walks, as shown in Figs. \ref{fig:eigenvalue_1step} and \ref{fig:eigenvalue_2step} respectively, the flow of eigenvalues in the complex plane as a function of the strength of nonlinearity is of
importance for the additional bifurcations of edge states. Here, we discuss further on these bifurcations, by pointing out that there is a fundamental difference for the stability analysis
between time-independent static systems and time-periodically driven systems, in which linearized equations are described by effective static Hamiltonians and time-evolution operators, respectively. On one hand, in the static systems, eigenvalues of non-Hermitian Hamiltonians such as $\omega$ in
Eq.\ (\ref{eq:continuum_limit_distribution}) are crucially important
for the stability of stationary states. On the other hand, in
the Floquet systems, eigenvalues of nonunitary time-evolution operators, such as $\lambda_1^{(\varepsilon,\eta)}$ and $\lambda_2^{(\varepsilon,\eta)}$ in Eqs.\ (\ref{eq:linearize_1step}) and (\ref{eq:linearize_2step}) respectively, determine the stability of stationary states. In the former case, a spectral region which makes a stationary state stable is unbounded as shown in Fig. \ref{fig:eigenvalue-motion} (a), while the stable region is bounded in the latter case, which is surrounded by the unstable region as shown in Fig. \ref{fig:eigenvalue-motion} (b).
This difference makes stationary states of Floquet systems more fragile. To clearly see this, we consider a simplified situation of the two-step nonlinear quantum walk in Sec.\
\ref{subsec:2step} by focusing on one eigenvalue, say $\lambda$, as shown by
the green triangle in Fig.\ \ref{fig:eigenvalue_2step}. Since $\lambda$ is real and monotonically decreases with increasing the strength of nonlinearity as shown in Fig. \ref{fig:eigenvalue-motion} (b), We can easily understand that $\lambda$ can flow into the unstable region in this case.
This peculiarity is more highlighted by introducing quasienergy defined from
$\lambda$ as $\xi = i \log (\lambda)$. The quasienergy $\xi$ plays a similar role of $\omega$, 
as the stability of stationary states depends on the sign of Im($\omega$) and Im($\xi$) 
as shown in Figs. \ref{fig:eigenvalue-motion} (a) and (c), respectively. Figure \ref{fig:eigenvalue-motion} (c) shows the trajectory of $\xi$ corresponding to $\lambda$ in Fig. \ref{fig:eigenvalue-motion} (b). When $\lambda$ passes through the origin, the value of $\xi$ jumps from $(0,-\infty)$ to $(\pi,-\infty)$ and the direction of motion is reversed, and then $\xi$ enters in the unstable region.
The discontinuity of $\xi$ originates from the phase singularity of $\lambda$ at the origin, and such a discontinuity never occurs in time-independent systems or systems described by effective static
Hamiltonians which are approximately derived. While we need to consider more complicated situations for the single-step nonlinear quantum walk, the boundedness of the stable region is important in both nonlinear quantum walks. Since the boundedness is a consequence of treating
time-evolution operators for linear stability analysis, the additional bifurcations of edge states that we have shown in the present work are phenomena unique to Floquet systems. It should be interesting to discuss the same type of bifurcations in which stationary states become unstable due to the boundedness of the stable region in other Floquet systems. To this end, we note that quantum walks are defined by simple time-evolution operators, and this advantage makes it possible to derive the bifurcation points analytically. Therefore, it may be reasonable to employ Floquet systems in which time-evolution operators can be derived without approximation, such as models treated in Refs. \cite{morimoto2017floquet,rudner2013anomalous}, while the derivation is difficult in Floquet systems described by Hamiltonians which are smooth functions of time, e.g. electronic systems under the application of light
\cite{oka2019floquet,oka2009photovoltaic,kitagawa2011transport}.
\\\indent
In the single-step nonlinear quantum walk, the effective Hamiltonian in the continuum limit is non-Hermitian (\ref{sec:continuum_limit}), although our analysis is based on time-evolution operators. Recently, systems described by non-Hermitian Hamiltonians with $\mathcal{PT}$ symmetry have been explored enthusiastically \cite{bender1998real,el2018non}, where $\mathcal{P}$ and $\mathcal{T}$ represent parity and time-reversal, respectively. As shown in \ref{sec:continuum_limit}, the effective Hamiltonian has $\mathcal{PT}$ symmetry. Although $\mathcal{PT}$ symmetry does not influence the stability of edge states in the single-step quantum walk (\ref{sec:continuum_limit}), generally, it can have the large effect on the stability of stationary states in nonlinear systems. While large number of $\mathcal{PT}$ symmetric non Hermitian systems are experimentally realized in linear optical systems, it may be also interesting to study the relation between stability of stationary states and $\mathcal{PT}$ symmetry in other nonlinear systems.\\\indent
An open problem is the stronger localization of probability distributions in the unstable region as shown in Fig.\ref{fig:time-evolution_1step} in Sec. \ref{sec:present}. Since the linear stability analysis is not useful in this region, other methods would lead to better understandings of stronger nonlinear effects in quantum walks.

 \section*{Acknowledgement}
We thank Y. Asano, H.\ Hirori, R.\ Okamoto, and K. Yakubo for helpful discussions.
This work was supported by KAKENHI (Grants No.\ JP18J20727, No.\ JP19H01838, No. JP18H01140, No.\ JP18K18733, and No.\ JP19K03646) and a Grant-in-Aid for Scientific Research on Innovative Areas (KAKENHI Grant No. JP15H05855 and No.\ JP18H04210) from the Japan Society for the Promotion of Science.

\appendix

\section{$\mathcal{PT}$ symmetry of the effective Hamiltonian in the continuum limit of the single-step nonlinear quantum walk}
\label{sec:continuum_limit}
We argue the non-Hermitian Hamiltonian in Eq.\ (\ref{eq:continuum_limit_distribution}) from the viewpoint of $\mathcal{PT}$ symmetry. $\mathcal{PT}$ symmetry is a combined symmetry of parity and
time-reversal symmetries. When a non-Hermitian operator has $\mathcal{PT}$ symmetry, there are two
phases, that is, a $\mathcal{PT}$-symmetry unbroken phase and
$\mathcal{PT}$-symmetry broken phase \cite{bender1998real,el2018non}. While all eigenvalues are real in
$\mathcal{PT}$-symmetry unbroken phase, in
the $\mathcal{PT}$-symmetry broken phase, eigenvalues are
partially or fully complex. The transition between the two
phases is called $\mathcal{PT}$-symmetry breaking.

Assuming a homogeneous system and applying Fourier transformation
to the infinitesimally weak fluctuating state $\ket{\delta
\psi_x(t)}$, Eq. (\ref{eq:continuum_limit_distribution}) is expressed
by a wave number $q$ as,
\begin{equation}
i\frac{\partial}{\partial t}\ket{\delta\psi_0(t)}=
\Omega(q)\ket{\delta\psi_0(t)},
\label{eq:continuum_limit_distribution_q}
\end{equation}
where $\Omega(q)$ corresponds to a non-Hermitian effective Hamiltonian
\begin{equation}
\hspace{-15mm}\Omega(q)=-2i\kappa\eta|\Phi|^2\identity_s+\tilde{\Omega}(q),\ 
\tilde{\Omega}(q)=\left(\begin{array}{cc}
q&-i\theta_0+2i\kappa|\Phi|^2\\
i\theta_0+2i\kappa|\Phi|^2&-q
\end{array}\right).
\label{eq:omega_q}
\end{equation}
Here, we focus only on the second term in Eq.\ (\ref{eq:omega_q}),
and identify that $\tilde{\Omega}(q)$ has $\mathcal{PT}$ symmetry
\begin{equation}
(\mathcal{PT})\tilde{\Omega}(q)(\mathcal{PT})^{-1}=\tilde{\Omega}(q),
\label{eq:pts}
\end{equation}
where the $\mathcal{PT}$ symmetry operator is
\begin{equation}
\mathcal{PT}=\sigma_3\mathcal{K}.
\label{eq:PT_operator}
\end{equation}

Since the eigenvalue of $\tilde{\Omega}(q)$ is
\begin{equation}
\tilde{\omega}=\pm\sqrt{q^2+\theta_0^2-4\kappa^2|\Phi|^4},
\label{eq:eigenvalue_omega_q}
\end{equation}
the eigenvalue of $\Omega(q)$ is given by
\begin{equation}
\omega=-2i\kappa \eta |\Phi|^2  \pm\sqrt{q^2+\theta_0^2-4\kappa^2|\Phi|^4}.
\label{eq:eigenvalue_omega}
\end{equation}
If the imaginary part of the eigenvalue of $\Omega(q)$ is negative (positive), the edge state
$\ket{\Phi_{\varepsilon,\eta}}$ is stable (unstable). The first term in Eq. (\ref{eq:omega_q}), $-2i\kappa\eta|\Phi|^2$, makes $\ket{\Phi_{\varepsilon,+}}$ ($\ket{\Phi_{\varepsilon,-}}$) stable
(unstable). When $\kappa<|\theta_0|/4|\Phi|^2$, $\tilde{\omega}$ is entirely real because the $\mathcal{PT}$ symmetry is unbroken. Therefore, the stability of the edge states is determined by the first pure imaginary term in Eq.\ (\ref{eq:eigenvalue_omega}).
Increasing $\kappa$, $\tilde{\omega}$ can become pure imaginary with positive and negative signs since the $\mathcal{PT}$ symmetry is broken when $\theta_0<2\kappa|\Phi|^2$. However,
the $\mathcal{PT}$ symmetry breaking does not affect the stability
of edge states since the first imaginary term in
Eq. (\ref{eq:eigenvalue_omega}) always dominates the sign of
${\rm Im}(\omega)$ due to the following relation
\begin{equation}
2\kappa|\Phi|^2\geq|{\rm Im}(\tilde{\omega})|.
\label{eq:magnitude}
\end{equation}
Therefore, the $\mathcal{PT}$ symmetry breaking is irrelevant to the stability of
$\ket{\Phi_{\varepsilon,\eta}}$ for the analysis in terms of the effective Hamiltonian.

\section{Derivation of the bifurcation points : upper bounds of the maximum
 for absolute values of eigenvalues for nonunitary operators}
\label{sec:bifurcation_points}
Given a square matrix $A$ which is diagonalizable, its spectral radius $\max(|\lambda_A|)$ and spectral norm $\sigma(A)$ always satisfy
\begin{equation}
\max(|\lambda_A|)\leq\sigma(A),\ \ 
\sigma(A)=\sqrt{\max(|\lambda_{A^\dagger A}|)},
\label{eq:inequality_sr-sn}
\end{equation}
where $\lambda_A$ ($\lambda_{A^\dagger A}$) is the eigenvalue of $A$
($A^\dagger A$). Here, $A^\dagger$ denotes Hermitian conjugation of $A$.
We can derive the bifurcation points in single and two-step
nonlinear quantum walks, where edge states become unstable, using Eq. (\ref{eq:inequality_sr-sn}).\\\indent
First, we derive the bifurcation points in the single-step
nonlinear quantum walk. As mentioned in the main text, the stability
of edge states $\ket{\Phi_{\varepsilon,\eta}}$ is determined by the maximum value of $|\lambda_1^{(\varepsilon,\eta)}|$, which is equivalent to the spectral radius of  $V_1^{(\varepsilon,\eta)}=U_1D_{\kappa,\eta}(|\Phi_x^{(\varepsilon)}|)$ in Eq.\ (\ref{eq:inequality_sr-sn}). Since $U_1$ is unitary, $V_1^{(\varepsilon,\eta)}$ satisfies
\begin{equation}
[V_1^{(\varepsilon,\eta)}]^\dagger
[V_1^{(\varepsilon,\eta)}]
=D^\dagger_{\kappa,\eta}(|\Phi_x^{(\varepsilon)}|)D_{\kappa,\eta}(|\Phi_x^{(\varepsilon)}|).
\label{eq:V1_daggaer_V1}
\end{equation}
Therefore, from Eq.\ (\ref{eq:inequality_sr-sn}), the upper limit
of $\max(|\lambda_1^{(\varepsilon,\eta)}|)$ is obtained by diagonalizing
$D^\dagger_{\kappa,\eta}(|\Phi_x^{(\varepsilon)}|)D_{\kappa,\eta}(|\Phi_x^{(\varepsilon)}|)$. Since $D_{\kappa,\eta}(|\Phi_x^{(\varepsilon)}|)$ defined in Eq.\ (\ref{eq:D_definition}) is Hermitian, the eigenvalues of $D^\dagger_{\kappa,\eta}(|\Phi_x^{(\varepsilon)}|)D_{\kappa,\eta}(|\Phi_x^{(\varepsilon)}|)$ correspond to $\delta_1^2$ and $\delta_2^2(x)$, where $\delta_1$ and $\delta_2(x)$ are the eigenvalues of $\mathcal{D}_{\kappa,\eta}(|\Phi_x^{(\varepsilon)}|)$, given by
\begin{equation}
\delta_1=1,\quad \delta_2(x) = 1-4\kappa\eta|\Phi_x^{(\varepsilon)}|^2.
\label{eq:eigenvalue_D_1step}
\end{equation}
From Eqs. (\ref{eq:inequality_sr-sn})-(\ref{eq:eigenvalue_D_1step}),
the spectral norm $\sigma[V_1^{(\varepsilon,\eta)}]$ is determined by
\begin{equation}
\sigma[V_1^{(\varepsilon,\eta)}] = \max_x[\delta_1,|\delta_2(x)|].
\end{equation}
On one hand, in the case of $\eta=-$, $1+4\kappa|\Phi_x^{(\varepsilon)}|^2$ becomes larger than $1$ and 
\begin{equation}
\max(|\lambda_1^{(\varepsilon,-)}|)\leq
1+4\kappa\max_x(|\Phi_x^{(\varepsilon)}|^2)\ (>1)
\label{eq:max_1step_-}
\end{equation}
is satisfied for nonzero $\kappa\ (>0)$. Thereby,
$\ket{\Phi_{\varepsilon,-}}$ can be always unstable.
On the other hand, in the case of $\eta=+$, the stability of
$\ket{\Phi_{\varepsilon,+}}$ depends on $\kappa$. When
$\max_x(|1-4\kappa|\Phi_x^{(\varepsilon)}|^2|)\leq1$ is
satisfied for small $\kappa$, $\ket{\Phi_{\varepsilon,+}}$ is stable
because of
\begin{equation}
\max(|\lambda_1^{(\varepsilon,+)}|)\leq1.
\label{eq:max_1step_+_1}
\end{equation}
When $1-4\kappa\max_x(|\Phi_x^{(\varepsilon)}|^2)<-1$ is satisfied,
$\ket{\Phi_{\varepsilon,+}}$ can be unstable due to
\begin{equation}
\max(|\lambda_1^{(\varepsilon,+)}|)\leq
|1-4\kappa\max(|\Phi_x^{(\varepsilon)}|^2)|\ (>1).
\label{eq:max_1step_+_2}
\end{equation}
The threshold for the stable to unstable transition is derived as
\begin{equation}
 \kappa_c = \frac{1}{2 \max_x(|\Phi_x^{(\varepsilon)}|^2)},
\end{equation}
from the condition that $1-4\kappa\max_x(|\Phi_x^{(\varepsilon)}|^2)=-1$.
When the value of $\max_x(|\Phi_x^{(\varepsilon)}|^2)$ is small, large $\kappa$ is needed to make $\ket{\Phi_{\varepsilon,+}}$ unstable. Therefore, $\ket{\Phi_{\varepsilon,+}}$ tends to be stable (unstable) if the localization length $[\log(1/\cos\theta_0+\tan\theta_0)]^{-1}$ is large (small).\\\indent
Second, we derive the bifurcation points in the two-step nonlinear
quantum walk, where the edge states $\ket{\Phi_{\varepsilon,\eta}}$ become unstable. In addition to Eq. (\ref{eq:inequality_sr-sn}), we also use an inequality for square matrices $B$ and $C$, 
\begin{equation}
\sigma(BC)\leq\sigma(B)\sigma(C).
\label{eq:inequality_sn-product}
\end{equation}
Substituting
$V_2^{(\varepsilon,\eta)}=U_{2b}D_{-\kappa,\tilde{\eta}}(|\tilde{\Phi}_x^{(\varepsilon)}|)U_{2a}D_{\kappa,\eta}(|\Phi_x^{(\varepsilon)}|)$
into $A$ in Eq. (\ref{eq:inequality_sr-sn}) and using Eq. (\ref{eq:inequality_sn-product}), we obtain 
\begin{equation}
\max(|\lambda_2^{(\varepsilon,\eta)}|)\leq
\sigma[D_{\kappa,\eta}(|\Phi_x^{(\varepsilon)}|)]\sigma[D_{-\kappa,\tilde{\eta}}(|\tilde{\Phi}_x^{(\varepsilon)}|)],
\label{eq:max_2step}
\end{equation}
since the spectral norms of the unitary operators, $U_{2a}$ and
$U_{2b}$, are always equal to $1$.
The eigenvalues of $D_{\kappa,\eta}(|\Phi_x^{(\varepsilon)}|)$ are given
in Eq.\ (\ref{eq:eigenvalue_D_1step}), and those of
$D_{-\kappa,\tilde{\eta}}(|\tilde{\Phi}_x^{(\varepsilon)}|)$ are given by
\begin{equation}
\tilde{\delta}_1=1,\quad \tilde{\delta}_2(x)=1+4\kappa\tilde{\eta}
|\tilde{\Phi}_x^{(\varepsilon)}|^2.
\label{eq:eigenvalue_D_2step_tilde}
\end{equation}
From Eqs.\ (\ref{eq:eigenvalue_D_1step}), (\ref{eq:max_2step}), and (\ref{eq:eigenvalue_D_2step_tilde}),
 the upper bound of $\max(|\lambda_2^{(\varepsilon,\eta)}|)$ is given by
\begin{equation}
\max(|\lambda_2^{(\varepsilon,\eta)}|) \le
 \max_x\left(\delta_1,|\delta_2(x)|\right)
 \max_x(\tilde{\delta}_1,|\tilde{\delta}_2(x)|).
\label{eq:max_2step_appendix}
\end{equation}
From Eq.\ (\ref{eq:max_2step_appendix}), we can understand that, for
 $(\varepsilon,\eta)=(0,+),\ (0,-),$ and $(\pi,-)$, 
 $\max_x(|\delta_2(x)|)$ and/or  $\max_x(|\tilde{\delta}_2(x)|)$ are larger than one for nonzero $\kappa\ (>0)$.
 Thereby, the  edge states $\ket{\Phi_x^{(0,\pm)}}$ and $\ket{\Phi_x^{(\pi,-)}}$ can be unstable. Regarding the edge state with $(\varepsilon,\eta)=(\pi,+)$, it is inevitably
stable as long as $\max_x(|\delta_2(x)|)\le 1$ and
 $\max_x(|\tilde{\delta}_2(x)|)\le 1$ are satisfied, which means
\begin{equation}
\max(|\lambda_2^{(\pi,+)}|)\leq1.
\label{eq:max_2step_pi_+_1}
\end{equation}
When one of $\max_x(|\delta_2(x)|)$ and $\max_x(|\tilde{\delta}_2(x)|)$ is larger than one, the upper bound of $\max(|\lambda_2^{(\pi,+)}|)$ becomes
\begin{equation}
\max(|\lambda_2^{(\pi,+)}|)\leq
\big|1-4\kappa\max_x(|\Phi_x^{(\pi)}|^2,
 |\tilde{\Phi}_x^{(\pi)}|^2)\big|
\label{eq:max_2step_pi_+_2}
\end{equation}
and $\ket{\Phi_{\pi,+}}$ can be unstable, since the right
hand side of Eq. (\ref{eq:max_2step_pi_+_2}) is larger than one. Therefore, the threshold of the stable to unstable transition is given by
\begin{equation}
 \kappa_c = \frac{1}{2\max_x(|\Phi_x^{(\pi)}|^2,
 |\tilde{\Phi}_x^{(\pi)}|^2)}.
\end{equation}

\section{Edge states in the single-step quantum walk}
\label{sec:edgestates_1step}
In \ref{sec:edgestates_1step}, we derive Eqs. (\ref{eq:edgestates_1step}) and (\ref{eq:gamma_N_1step}). Since edge states satisfy Eq. (\ref{eq:LR_edgestates}), in the inner region where $-m\leq x \leq m$ and $\theta_0(x)=\theta_0$, the time-evolution equation of $\ket{\Phi_{\varepsilon,\eta}}$ becomes
\begin{equation}
e^{-i\varepsilon}\Phi_{x,L}^{(\varepsilon,\eta)}
=\cos(\theta_0/2)\alpha_\eta\Phi_{x+1,L}^{(\varepsilon,\eta)}
-\sin(\theta_0/2)\beta_\eta\Phi_{x-1,L}^{(\varepsilon,\eta)},
\label{eq:Phi_L_1step}
\end{equation}
\begin{equation}
\eta e^{-i\varepsilon}\Phi_{x,R}^{(\varepsilon,\eta)}
=\sin(\theta_0/2)\alpha_\eta\Phi_{x+1,R}^{(\varepsilon,\eta)}
+\cos(\theta_0/2)\beta_\eta
\Phi_{x-1,R}^{(\varepsilon,\eta)},
\label{eq:Phi_R_1step}
\end{equation}
where $\alpha_\eta$ and $\beta_\eta$ are
\begin{equation}
\alpha_\eta=\cos(\theta_0/2)-\eta\sin(\theta_0/2),\ \ 
\beta_\eta=\sin(\theta_0/2)+\eta\cos(\theta_0/2).
\label{eq:alpha-beta_1step}
\end{equation}
Considering edge states localized near the right boundary, we assume
\begin{equation}
\Phi_{x,s}^{(\varepsilon,\eta)}
=N_1(-1)^{\frac{\varepsilon}{\pi}x} e^{-\gamma(m-x)},
\label{eq:hypothesis_1step}
\end{equation}
where $N_1$ is the normalization constant. While $\Phi_{x,s}^{(0,\eta)}$ have the same sign in any position, the sign of $\Phi_{x,s}^{(\pi,\eta)}$ is opposite to that of $\Phi_{x\pm1,s}^{(\pi,\eta)}$. Substituting Eq. (\ref{eq:hypothesis_1step}) into Eqs. (\ref{eq:Phi_L_1step}) and (\ref{eq:Phi_R_1step}), we obtain
\begin{equation}
1=\cos(\theta_0/2)\alpha_\eta e^\gamma
-\sin(\theta_0/2)\beta_\eta
e^{-\gamma},
\label{eq:gamma_L_1step}
\end{equation}
\begin{equation}
\eta=\sin(\theta_0/2)\alpha_\eta e^\gamma
+\cos(\theta_0/2)\beta_\eta e^{-\gamma},
\label{eq:gamma_R_1step}
\end{equation}
for both $\varepsilon=0$ and $\varepsilon=\pi$. Since Eqs. (\ref{eq:gamma_L_1step}) and (\ref{eq:gamma_R_1step}) are quadratic equations of $e^\gamma$, we obtain two solutions of $e^\gamma$ from each equation. We employ one solution which is obtained from both Eqs. (\ref{eq:gamma_L_1step}) and (\ref{eq:gamma_R_1step}). Then, $e^\gamma$ becomes
\begin{equation}
e^\gamma=\beta_\eta/\alpha_\eta.
\label{eq:gamma}
\end{equation}
From Eqs. (\ref{eq:alpha-beta_1step}) and
(\ref{eq:gamma}), $\eta$ must be $+$ for the solution of the
localized state  ($\gamma>0$). In the outer region, where $x\geq m+1$
and $\theta_0(x)=-\theta_0$, we obtain the same solution in the same
way, while Eq. (\ref{eq:hypothesis_1step}) is changed to
$\Phi_{x,s}^{(\varepsilon,\eta)}=N_1(-1)^{\frac{\varepsilon}{\pi}x}e^{-\gamma[x-(m+1)]}$. In
the case $\ket{\Phi_{\varepsilon,\eta}}$ is localized near the left
boundary, $\eta$ must be $-$, while the localization length $\gamma^{-1}$ is the same. The boundary condition of $\ket{\Phi_{\varepsilon,+}}$ is
\begin{equation}
e^{-i\varepsilon}
\Phi_{m,L}^{(\varepsilon,+)}=\alpha_-[\cos(\theta_0/2)
\Phi_{m+1,L}^{(\varepsilon,+)}
-\sin(\theta_0/2)e^{-\gamma}
(-1)^\frac{\varepsilon}{\pi}\Phi_{m,L}^{(\varepsilon,+)}],
\label{eq:boundary_condition_L_1step}
\end{equation}
\begin{equation}
e^{-i\varepsilon}
\Phi_{m,R}^{(\varepsilon,+)}=\beta_+[\sin(\theta_0/2)
\Phi_{m+1,R}^{(\varepsilon,+)}
+\cos(\theta_0/2)e^{-\gamma}
(-1)^\frac{\varepsilon}{\pi}\Phi_{m,R}^{(\varepsilon,+)}],
\label{eq:boundary_condition_R_1step}
\end{equation}
near the right boundary. Assuming
\begin{equation}
\Phi_{m,s}^{(\varepsilon,+)}=
(-1)^\frac{\varepsilon}{\pi}\Phi_{m+1,s}^{(\varepsilon,+)},
\label{eq:bondary_condition_1step}
\end{equation}
eqs. (\ref{eq:boundary_condition_L_1step}) and (\ref{eq:boundary_condition_R_1step}) are satisfied. For edge states with $\eta=-$ and localized near the left boundary, the boundary condition is the same with Eq. (\ref{eq:bondary_condition_1step}), changing $m$, $m+1$, and $+$ into $-m$, $-m-1$, and $-$, respectively.

\section{Edge states in the two-step quantum walk}
\label{sec:edgestates_2step}
In \ref{sec:edgestates_2step}, we derive
Eqs. (\ref{eq:edgestates_2step_0_+})-(\ref{eq:gamma_pi_B}). The edge states have amplitudes only in even sites or odd sites, since wave funtions at even (odd) sites always shift to even (odd) sites after one time step, in the two-step quantum walk described by $U_2$. Although we derive edge states at even sites, edge states in odd sites are obtained in the same way. In the inner region where $-m\leq x \leq m$ and $\theta_i(x)=\theta_i$ ($i=1,2$), Eq. (\ref{eq:varepsilon_eta}) can be wriiten as
\begin{equation}
\hspace{-15mm}e^{-i\varepsilon}\Phi_{x,L}^{(\varepsilon,\eta)}=\begin{array}{l}
\cos(\theta_1/2)\cos(\theta_2)\alpha_{1,\eta}
\Phi_{x+2,L}^{(\varepsilon,\eta)}
-\cos(\theta_1/2)\sin(\theta_2)\beta_{1,\eta}
\Phi_{x,L}^{(\varepsilon,\eta)}\\
-\sin(\theta_1/2)\sin(\theta_2)\alpha_{1,\eta}
\Phi_{x,L}^{(\varepsilon,\eta)}
-\sin(\theta_1/2)\cos(\theta_2)\beta_{1,\eta}
\Phi_{x-2,L}^{(\varepsilon,\eta)}
\end{array}
\label{eq:phi_L_2step}
\end{equation}
\begin{equation}
\hspace{-15mm}\eta e^{-i\varepsilon}\Phi_{x,R}^{(\varepsilon,\eta)}=\begin{array}{l}
\sin(\theta_1/2)\cos(\theta_2)\alpha_{1,\eta}
\Phi_{x+2,R}^{(\varepsilon,\eta)}
-\sin(\theta_1/2)\sin(\theta_2)\beta_{1,\eta}
\Phi_{x,R}^{(\varepsilon,\eta)}\\
+\cos(\theta_1/2)\sin(\theta_2)\alpha_{1,\eta}
\Phi_{x,R}^{(\varepsilon,\eta)}
+\cos(\theta_1/2)\cos(\theta_2)\beta_{1,\eta}
\Phi_{x-2,R}^{(\varepsilon,\eta)}
\end{array}
\label{eq:phi_R_2step}
\end{equation}
where $\alpha_{i,\eta}$ and $\beta_{i,\eta}$ are defined as
\begin{equation}
\alpha_{i,\eta}=\cos(\theta_i/2)-\eta\sin(\theta_i/2),\ 
\beta_{i,\eta}=\sin(\theta_i/2)+\eta\cos(\theta_i/2).
\label{eq:alpha-beta_2step}
\end{equation}
Considering the edge states localized near the left boundary, we assume
\begin{equation}
\Phi_{x,s}^{(\varepsilon,\eta)}
=N_2(-1)^\frac{\varepsilon x}{2\pi}e^{-\gamma_\varepsilon(x+m)}, 
\label{eq:hypothesis_2step}
\end{equation}
where $N_2$ is the normalization constant. Substituting Eq. (\ref{eq:hypothesis_2step}) into Eqs. (\ref{eq:phi_L_2step}) and (\ref{eq:phi_R_2step}), we obtain
\begin{equation}
1=\begin{array}{l}
\cos(\theta_1/2)
[\cos(\theta_2)\alpha_{1,\eta} e^{-2\gamma_\varepsilon}
-(-1)^\frac{\varepsilon}{\pi}\sin(\theta_2)\beta_{1,\eta}]\\
-\sin(\theta_1/2)
[(-1)^\frac{\varepsilon}{\pi}\sin(\theta_2)\alpha_{1,\eta}
+\cos(\theta_2)\beta_{1,\eta} e^{2\gamma_\varepsilon}]
\end{array}
\label{eq:gamma_L_2step}
\end{equation}
\begin{equation}
\eta=\begin{array}{l}
\sin(\theta_1/2)
[\cos(\theta_2)\alpha_{1,\eta} e^{-2\gamma_\varepsilon}
-(-1)^\frac{\varepsilon}{\pi}\sin(\theta_2)\beta_{1,\eta}]\\
+\cos(\theta_1/2)
[(-1)^\frac{\varepsilon}{\pi}\sin(\theta_2)\alpha_{1,\eta}
+\cos(\theta_2)\beta_{1,\eta} e^{2\gamma_\varepsilon}]
\end{array}
\label{eq:gamma_R_2step}
\end{equation}
Solving Eqs. (\ref{eq:gamma_L_2step}) and (\ref{eq:gamma_R_2step}) for $e^{2\gamma_\varepsilon}$, we obtain four solutions. Among them, we choose one which satisfies both Eqs. (\ref{eq:gamma_L_2step}) and (\ref{eq:gamma_R_2step}). Then, $\gamma_0$ and $\gamma_\pi$ are
\begin{equation}
e^{2\gamma_0}=(\alpha_{1,\eta}/\beta_{1,\eta})
(\alpha_{2,\eta}/\beta_{2,\eta}),\ 
e^{2\gamma_\pi}=(\alpha_{1,\eta}/\beta_{1,\eta})
(\beta_{2,\eta}/\alpha_{2,\eta}),
\label{eq:gamma_2step}
\end{equation}
respectively. In the outer region where $x\leq-m-1$ and $\theta_i(x)=-\theta_i$, $\gamma_0$ and $\gamma_\pi$ become the same with Eq. (\ref{eq:gamma_2step}), while Eq. (\ref{eq:hypothesis_2step}) is changed to $\Phi_{x,s}^{(\varepsilon,\eta)}=N_2(-1)^\frac{\varepsilon x}{2\pi}e^{-\gamma_\varepsilon(-m-2-x)}$. As mentioned in the main text, we focus only on parameter regions A and B in Fig. \ref{fig:theta12}, for simplicity. From Eqs. (\ref{eq:alpha-beta_2step}) and (\ref{eq:gamma_2step}), it is known that chirality must be minus for $\varepsilon=0$, as $\gamma_0>0$. For $\varepsilon=\pi$, $\gamma_\pi>0$ is satisfied when $\eta=+$, in the parameter region A where $0<\theta_2<\pi/2$ and $-\theta_2<\theta_1<\theta_2$. In the parameter region B where $0<\theta_1<\pi/2$ and $-\theta_1<\theta_2<\theta_1$, $\eta=-$ for $\gamma_\pi$ to be positive. Edge states localized near the right boundary have the opposite chirality, while $\gamma_0$ and $\gamma_\pi$ have the same values with Eq. \ref{eq:gamma_2step}). Near the left boundary, $\Phi_{-m,s}^{(\varepsilon,\eta)}$ satisfies
\begin{equation}
\hspace{-8mm}-\sin(\theta_1/2)\cos(\theta_2)\eta\alpha_{1,\eta}
\Phi_{-m-2,s}^{(\varepsilon,\eta)}=\begin{array}{l}
[1+\eta(-1)^\frac{\varepsilon}{\pi}\sin(\theta_2)]
\Phi_{-m,s}^{(\varepsilon,\eta)}\\
-\cos(\theta_1/2)\cos(\theta_2)\alpha_{1,\eta} 
\Phi_{-m+2,s}^{(\varepsilon,\eta)}
\end{array}
\label{eq:boundary_condition_2step}
\end{equation}
Substituting Eqs. (\ref{eq:hypothesis_2step}) and (\ref{eq:gamma_2step}) into Eq. (\ref{eq:boundary_condition_2step}), the boundary condition of $\Phi_{x,s}^{(0,-)}$ is
\begin{equation}
\cos(\theta_2)\Phi_{-m-2,s}^{(0,-)}
=[1-\sin(\theta_2)]\Phi_{-m,s}^{(0,-)},
\label{eq:boundary_condition_2step_0}
\end{equation}
in both parameter regions. For $\ket{\Phi_{\pi,\eta}}$, taking it into
account the dependence of $\eta$ and $\gamma_\pi$ on parameter regions, boundary conditions become
\begin{equation}
\cos(\theta_2)\Phi_{-m-2,s}^{(\pi,+)}
=[1-\sin(\theta_2)]\Phi_{-m,s}^{(\pi,+)}
\label{eq:boundary_condition_2step_pi_A}
\end{equation}
in the parameter region A, and
\begin{equation}
[1-\sin(\theta_2)]\Phi_{-m-2,s}^{(\pi,-)}
=\cos(\theta_2)\Phi_{-m,s}^{(\pi,-)}
\label{eq:boundary_condition_2step_pi_B}
\end{equation}
in the parameter region B, from Eqs. (\ref{eq:hypothesis_2step}), (\ref{eq:gamma_2step}), and (\ref{eq:boundary_condition_2step}). The boundary conditions near the right boundary are obtained in the same way.

\section{Derivation of Eq. (\ref{eq:chirality_Phi_tilde})}
\label{sec:chirality_2step}
In order to derive Eq. (\ref{eq:chirality_Phi_tilde}), we define two time evolution operators with chiral symmetry $\Gamma U \Gamma^{-1}=U^{-1}$, $U=U^\prime$ and $U=U^{\prime\prime}$. In the symmtery time frame \cite{asboth2012symmetries}, they are defined as
\begin{equation}
U^\prime=U_bU_a,\ U^{\prime\prime}=U_aU_b,
\label{eq:Up_Upp}
\end{equation}
where $U_a$ and $U_b$ satisfy
\begin{equation}
\Gamma U_a \Gamma^{-1}=U_b^{-1}.
\label{eq:chiral_symmetry}
\end{equation}
Note that, in the main text, $U_a$, $U_b$, and $U^\prime$ correspond to
$U_{2a}$, $U_{2b}$, and $U_2$, respectively. The edge states in the system described by $U^\prime$ are distinguished by the quasienergy $\varepsilon=0,\pi$ 
\begin{equation}
U^\prime\ket{\Phi^\prime_{\varepsilon,\eta}}
=e^{-i\varepsilon}\ket{\Phi^\prime_{\varepsilon,\eta}},
\label{eq:quasi_energy_Up}
\end{equation}
and chirality $\eta=\pm$,
\begin{equation}
\Gamma\ket{\Phi^\prime_{\varepsilon,\eta}}
=\eta\ket{\Phi^\prime_{\varepsilon,\eta}}.
\label{eq:chirality_Up}
\end{equation}
Acting $U_a$ from left on both sides of Eq. (\ref{eq:quasi_energy_Up}) and taking Eq. (\ref{eq:Up_Upp}) into account, we obtain
\begin{equation}
U^{\prime\prime}U_a\ket{\Phi^\prime_{\varepsilon,\eta}}
=e^{-i\varepsilon}U_a\ket{\Phi^\prime_{\varepsilon,\eta}}.
\label{eq:aba}
\end{equation}
From Eq. (\ref{eq:aba}), it is seen that edge states with quasienergy $\varepsilon$ in the system described by $U^{\prime\prime}$, $\ket{\Phi_{\varepsilon,\tilde{\eta}}^{\prime\prime}}$, are obtained from $\ket{\Phi^\prime_{\varepsilon,\eta}}$,
\begin{equation}
\ket{\Phi_{\varepsilon,\tilde{\eta}}^{\prime\prime}}
=U_a\ket{\Phi^\prime_{\varepsilon,\eta}},
\label{eq:eigenstates_pp_p}
\end{equation}
ignoring a global phase. Chirality of $\ket{\Phi_{\varepsilon,\tilde{\eta}}^{\prime\prime}},\ \tilde{\eta},$ is determined by $\varepsilon$ and chirality of $\ket{\Phi^\prime_{\varepsilon,\eta}}$, $\eta$. Acting the chiral symmetry operator $\Gamma$ from left on both sides of Eq. (\ref{eq:eigenstates_pp_p}), $\Gamma\ket{\Phi_{\varepsilon,\tilde{\eta}}^{\prime\prime}}$ becomes
\begin{equation}
\Gamma\ket{\Phi_{\varepsilon,\tilde{\eta}}^{\prime\prime}}
=U_b^{-1}\eta\ket{\Phi^\prime_{\varepsilon,\eta}},
\label{eq:transform_1}
\end{equation}
since Eqs. (\ref{eq:chiral_symmetry}) and (\ref{eq:chirality_Up}) hold. Using Eqs. (\ref{eq:Up_Upp}) and (\ref{eq:quasi_energy_Up}), the right hand side of Eq. (\ref{eq:transform_1}) can be written as
\begin{equation}
U_b^{-1}\eta\ket{\Phi^\prime_{\varepsilon,\eta}}=
U_b^{-1}\eta e^{i\varepsilon}
U_bU_a\ket{\Phi^\prime_{\varepsilon,\eta}}.
\label{eq:transform_2}
\end{equation}
From Eqs. (\ref{eq:eigenstates_pp_p}), (\ref{eq:transform_1}), and (\ref{eq:transform_2}), $\ket{\Phi_{\varepsilon,\tilde{\eta}}^{\prime\prime}}$ satisfies
\begin{equation}
\Gamma\ket{\Phi_{\varepsilon,\tilde{\eta}}^{\prime\prime}}
=\tilde{\eta}\ket{\Phi_{\varepsilon,\tilde{\eta}}^{\prime\prime}},\ 
\tilde{\eta}=\eta e^{i\varepsilon},
\label{eq:chirality_pp}
\end{equation}
which means that Eq. (\ref{eq:chirality_Phi_tilde}) is satisfied, where $\ket{\Phi^\prime_{\varepsilon,\eta}}$ and $\ket{\Phi_{\varepsilon,\tilde{\eta}}^{\prime\prime}}=U_a\ket{\Phi^\prime_{\varepsilon,\eta}}$ correspond to $\ket{\Phi_{\varepsilon,\eta}}$ and $\ket{\tilde{\Phi}_{\varepsilon,\eta}}$, respectively, in the main text. Therefore, chirality of $\ket{\Phi_{0,\tilde{\eta}}^{\prime\prime}}$ is the same as that of $\ket{\Phi^\prime_{0,\eta}}$, $\eta$. On the other hand, $\ket{\Phi_{\pi,\tilde{\eta}}^{\prime\prime}}$ has chirality opposite to that of $\ket{\Phi^\prime_{\pi,\eta}}$, $-\eta$.

 \section*{Reference}
\bibliographystyle{iopart-num}
\bibliography{reference.bib}

\end{document}